\documentclass[journal]{IEEEtran}
\usepackage{cite}
\usepackage{array} 
\usepackage{multirow}
\usepackage{amsmath,amssymb,amsfonts}
\usepackage{algorithmic}
\usepackage{graphicx}
\usepackage{textcomp}
\usepackage{xcolor}
\usepackage{pgfplots}
\usepackage[keeplastbox]{flushend}

\def\BibTeX{{\rm B\kern-.05em{\sc i\kern-.025em b}\kern-.08em
    T\kern-.1667em\lower.7ex\hbox{E}\kern-.125emX}}
\usepackage{soul}

\begin{document}

\title{Aspects of Quality in Internet of Things (IoT) Solutions: A Systematic Mapping Study}

\author{Bestoun S. Ahmed*,
        Miroslav Bures,
        Karel Frajtak, and
        Tomas Cerny
        
\thanks{B. Ahmed is with the Department of Computer Science, Faculty of Electrical Engineering, Czech Technical University in Prague, Czech Republic, and the Department of Mathematics and Computer Science, Karlstads University, Karlstad, Sweden
email: albeybes@fel.cvut.cz}
\thanks{M. Bures and K. Frajtak are with the Department of Computer Science, Faculty of Electrical Engineering, Czech Technical University, Karlovo nam. 13, Prague, Czech Republic
email: buresm3@fel.cvut.cz}
\thanks{T. Cerny is with the Dept. of Computer Science, ECS, Baylor University, One Bear
Place \#97141, Waco, TX, 76798, USA, email: tomas\_cerny@baylor.edu}}

\maketitle

\begin{abstract}
Internet of Things (IoT) is an emerging technology that has the promising power to change our future. Due to the market pressure, IoT systems may be released without sufficient testing. However, it is no longer acceptable to release IoT systems to the market without assuring the quality. As in the case of new technologies, the quality assurance process is a challenging task. This paper shows the results of the first comprehensive and systematic mapping study to structure and categories the research evidence in the literature starting in 2009 when the early publication of IoT papers for IoT quality assurance appeared. The conducted research is based on the most recent guidelines on how to perform systematic mapping studies. A set of research questions is defined carefully regarding the quality aspects of the IoT. Based on these questions, a large number of evidence and research papers is considered in the study (478 papers). We have extracted and analyzed different levels of information from those considered papers. Also, we have classified the topics addressed in those papers into categories based on the quality aspects. The study results carry out different areas that require more work and investigation in the context of IoT quality assurance. The results of the study can help in a further understanding of the research gaps. Moreover, the results show a roadmap for future research directions.
\end{abstract}


\IEEEpeerreviewmaketitle

\section{Introduction}

\IEEEPARstart{I}{nternet} of things (IoT) is an evolving technological topic that gained importance recently due to its potential impact on our daily life and future societies. It is expected that in the near future our cars, consumer products, industries, and other everyday objects become collaborative via an Internet connection and robust data analysis. This combination of connective objects with data analysis capabilities could be an authoritative source for the intelligent decisions that could transform the way people live in the future. In 1999, the British technology pioneer Kevin Ashton introduced the term ``Internet of Things'' (IoT) to describe the ability of connected sensors on the Internet to bring new services\cite{Kavin2009}. Although the term was new, the concept has been around for decades when computer and networks combined to control and monitor devices.

\begin{figure*}[h]
\centering \includegraphics[scale=0.37]{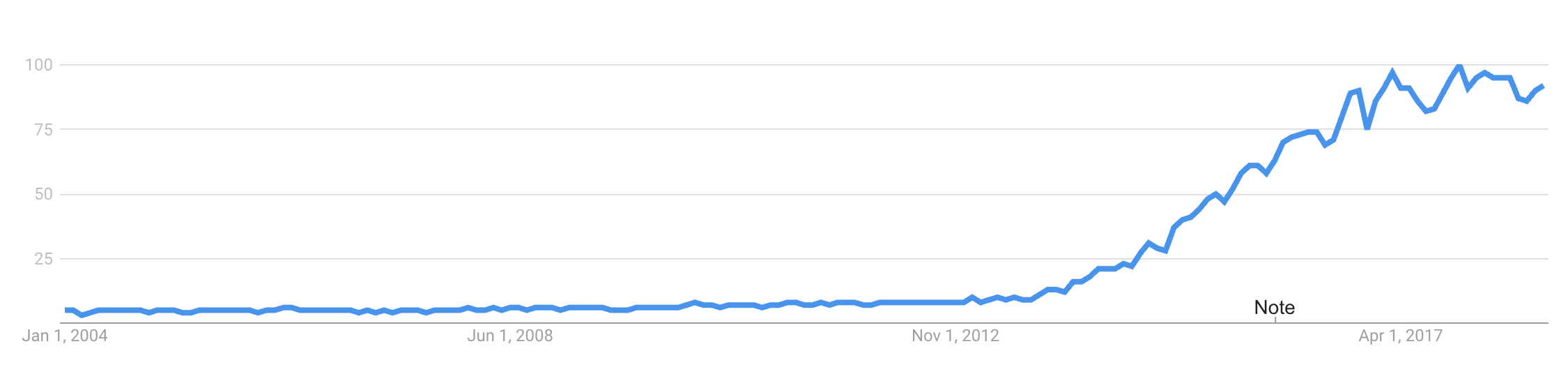} 
\vspace{-1.3em}\caption{Google search trends since 2004 for terms Internet of Things and IoT. The relative value 0 to 100 represents search interest.}
\label{fig:GoogleTrend} 
\vspace{-.7em}
\end{figure*}

\begin{figure*}[h]
\centering \includegraphics[width=5.3in]{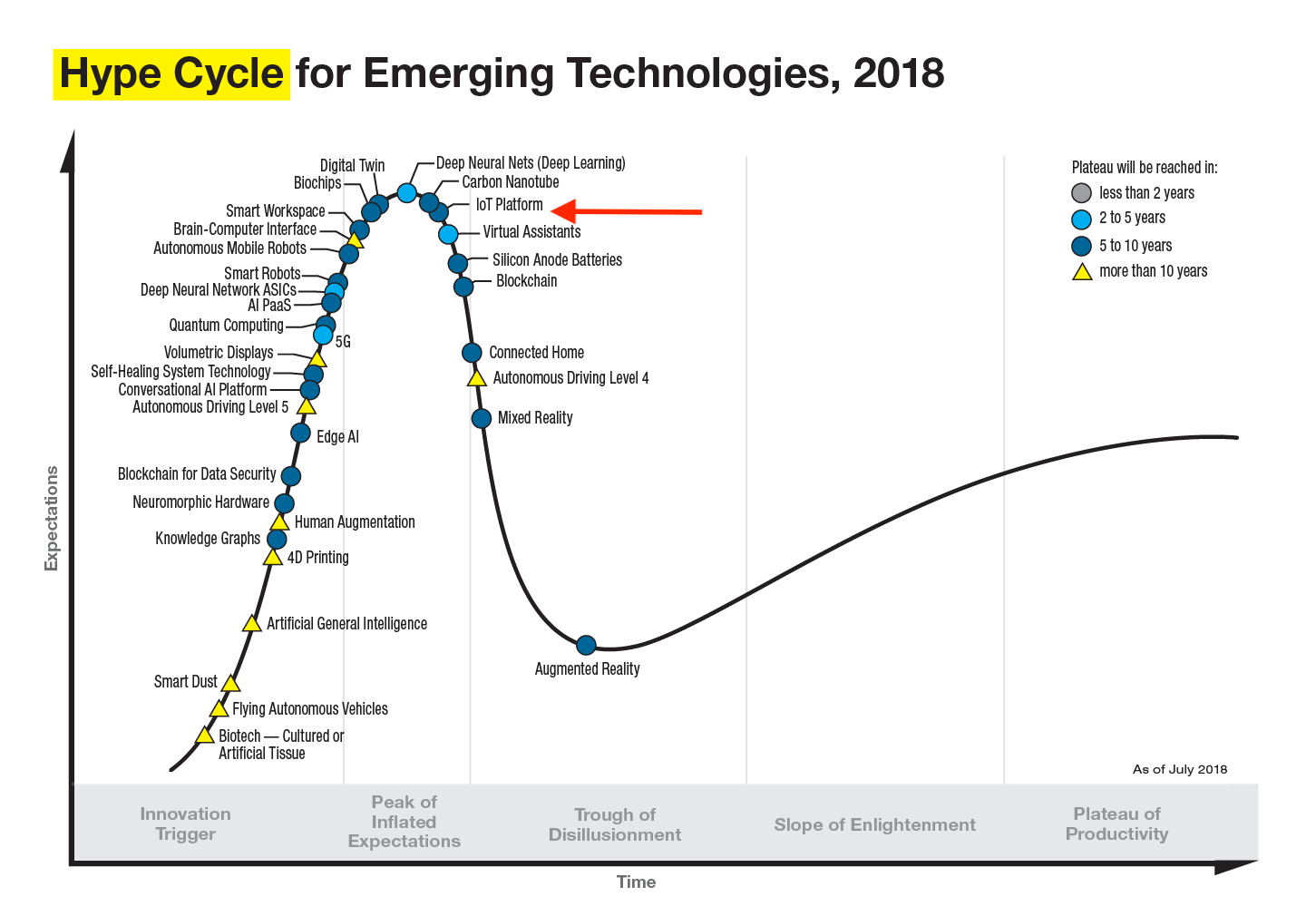}
\vspace{-1.3em}
\caption{Gartner 2018 Hype Cycle of emerging technologies. Source Gartner Inc.
\cite{GartnerHype}}
\label{fig:Gartner} 
\vspace{-1em}
\end{figure*}

The applications using the concepts of IoT are emerging tremendously day by day. These applications span a broad range of eHealth, security, entertainment, smart cities, defense and in many other necessary directions. Also, more objects are going to gain the ability of direct Internet connection in the coming years. As this range of connectivity and application expands, the IoT is going to affect our personal lives and public safety directly. With its expansion, the chance of system and network failure in IoT becomes higher than before. To this end, it is not acceptable to develop poor quality IoT systems that might cause loss of fortune, data or even life. For this reason, Quality Assurance (QA) is an essential and valuable issue for the IoT systems before the consolidation of these sensors, devices, applications, and systems go to the market. For example, to ensure accurate delivery timing, a shipment tracking system may use many sensors that communicate with many back-end software and many sophisticated algorithms. This system needs a robust QA process to validate the algorithms and workflow carefully.

In conventional software engineering process, the QA spans each step of the software development process, which aims to deliver software with minimum defects, meeting specific levels of functionality according to reliability and performance \cite{Maxim2016}. As with the case of an evolving technology, the QA aspect of IoT must be defined clearly and should be optimized. These QA aspects must also be improved periodically to meet the market and user expectations of many core principles, such as security, privacy, compatibility, reliability, and many others aspects. Moreover, it is the systematic pattern of all actions for providing and proving the ability of a software process to build high--quality products \cite{Linda2002}. It also attempts to improve the development process from the requirement step till the end \cite{Shoemaker2016}. Thus, it improves software functionality, including safety and reliability. In fact, the QA for IoT does not deviate from this context. Ensuring the quality of IoT system is a more challenging process than ensuring the quality of software thanks to the interaction of different objects, usage of various platforms, configurations, and input domains. Ensuring the quality of this type of system requires a framework to evaluate each component individually and verify the expected output.

This paper provides the first comprehensive and systematic mapping study to organize and categorize the research evidence in the literature starting in 2009 when the early publication of the QA aspect of IoT systems papers started. The goal of the study is to identify the quantity, the results available, type of research and disclose available research opportunities for the future. The study also tries to answer important research questions in the context of QA for IoT. The most up--to--date methodologies and guidelines are used in the study to collect and analyze the related studies published. In doing so, methods and aspects of quality assurance are addressed in the study as there is no comprehensive and dedicated study in this direction. The study selects the evidence and papers published in the past starting from 2009. The paper aims to serve as a guide for future researchers by providing an unbiased mapping study of the published research and addressing several significant research questions (RQ). 

The rest of this paper is organized as follows: Section \ref{RelatedWork} presents the motivation and the overview of related work for this study. Section \ref{Method} describes the methodology of the mapping study. Section \ref{sec:Results} presents the results and outcomes of the study followed by a discussion in Section \ref{sec:Discussion}. Threats to the validity of the study are given in Section \ref{sec:Threats-to-Validity}. Finally, Section \ref{sec:Conclusions} concludes the work.

\section{Motivation and Related Works}

\label{RelatedWork}

Since its existence, IoT gained more popularity, especially in the last decade. For example, Figure \ref{fig:GoogleTrend} shows the web search measured by Google search trends for \textquotedblleft IoT\textquotedblright{} and \textquotedblleft Internet of Things\textquotedblright{} words since the first search event until September 2018. The figure shows us the increasing interest in IoT owing to its influence and impact. This influence and impact on our life are expected to increase in the coming years. For example, Garter\textquoteright s Information Technology Hype Cycle \cite{GartnerHype} identified IoT as one of the emerging technologies in the coming decade as of August 2018. Figure \ref{fig:Gartner} shows the expected time periods for the emerging technologies that have been identified by the Hype Cycle. Garter's Information Technology expects that IoT platforms will take around 5-10 years for entirely market adoption (See the red arrow in Figure \ref{fig:Gartner}).

Today's impact of IoT and its expectation for the future is beyond the term. A number of expectations about the potential impact of IoT on the economy for coming years are also available. For example, Cisco expects more than 24 billion connected objects to the Internet by 2019 \cite{IoTTechnicalReport2015}, Morgan Stanley expects more than 75 billion by 2020 \cite{IoTTechnicalReport2015}, and Huawei expects 100 billion by 2025 \cite{Huawei}. Of course, the financial impact of these connected objects on the future global economy is enormous. For example, McKinsey Global Institute expected this impact as much as 3.9 to 11.1 trillion US dollars by 2025 \cite{McKinsey}. While these numbers are expectations, however, they can tell us the potential impact of IoT on the future.

As a reflection of this importance and expectations, many research groups are working on projects related directly or indirectly to IoT. As a result, many papers coming out from these research activities. To summarize these activities, collect these efforts, and identify research directions, few survey and mapping study papers tried to address IoT. However, due to the broad area of IoT's research, it is impossible to address those research activities and evidence in one study. To this end, researchers tried to do several survey and mapping studies for specific areas in IoT. However, these studies do not directly focus on quality aspects of the contemporary IoT systems and primarily does not discuss quality assurance and testing techniques in this context.

For instance, Stojkoska and Trivodaliev \cite{Stojkoska2017} performed a systematic study for the state--of--the--art applications for smart homes to identify the challenges and solutions for integration into IoT environments. In these challenges, two QA--related aspects are discussed, namely interoperability and security and privacy. However, a discussion of quality aspects is not the primary goal of this study and these two issues are not analyzed to a broader extent. Also, another QA--related issues as challenged testing of the IoT system caused by a number of possible configurations, backward--compatibility testing issues or reliability of the system are not targeted in the study. 

Ray \cite{Ray2016} performed a systematic review study to survey the existing architectures of IoT applications that are solving real--life problems. Unfortunately, implications of the particular architectures for the quality of the solution is not analyzed systematically in this study. Ray discusses the reliability of the system on several occasions in the paper. Moreover, this discussion is not conducted in a focused and systematic manner, as it was not the main goal of the paper.

Atzori \textit{et al.} \cite{Atzori2010} performed a systematic literature study on IoT from the communication and network perspective to show different implemented network paradigms, communication protocols, wired and wireless sensors used in the literature. This paper summarized the IoT--related state--of--the--art in the year 2010 and also mentions security and privacy concerns. Nevertheless, the paper does not discuss all other QA--related issues of IoT systems, partially because of its publication date and because it was not the primary focus of the study.

Perera \textit{et al.} \cite{Perera2014} performed a survey study to address the context--awareness issues in the literature from an IoT perspective. The paper discussed Quality of Context (QoC), having an impact on the quality of data processed by the IoT system. Further, security and privacy aspects of the IoT systems are discussed as the significant concern in context--aware systems. Possibilities of quality checking are also analyzed during categorization of context reasoning decision modeling techniques discussed in the paper. All these quality--related discussions are, however, driven by the primary analysis of the context--awareness IoT concepts and the discussion does not extend to other relevant quality--related areas.

Whitmore \textit{et al.}\cite{Whitmore2015} performed a study to report the state--of--the--art on IoT to identify the future research directions. The study looks into the general research direction rather than a particular area of research in IoT to identify the open research questions. Despite its broad focus, the study does not analyze QA--related issues and challenges of the IoT systems; in the challenges discussed in the study, security, and privacy are mentioned briefly. The study classifies the papers to three--level of categories. However, quality, testing, and verification of IoT systems have not been dedicated to any category in this study. Considering the fact that the paper is published in 2015, this is an implication of the design of the study and composition of search strings, as the significant number of IoT quality--related papers have been published from 2012, as we show in Figure \ref{fig:Publication-per-year} later on.

Gubbi \textit{et al.} \cite{Gubbi2013} performed a literature study to clarify the elements and architectures used for IoT in the literature. In the open challenges and future directions related to IoT systems quality aspects, Gubbi \textit{et al.} mention energy efficiency, security, protocols, and Quality of Service (QoS). From these aspects, only QoS and security are given broader floor in the analysis of the state--of--the--art. 

QoS in IoT has been addressed by a separate systematic mapping study by White \textit{et al.} \cite{WHITE2017186}. Several quality models are surveyed and discussed and software product quality model ISO/IEC 25010 is discussed in this context. This systematic study focuses only on rather a narrow scope of QA when compared to the scope of this paper. The study by White \textit{et al.} primarily analyzes papers dedicated to QoS, quality models, monitoring and Service Level Agreement, as the authors clearly state in the description of search methodology. 

In all the studies mentioned above, in addition to the state--of--the--art research addressing, the research challenges have also been discussed along with the future directions for research.

Several systematic literature reviews were also conducted in the areas, which closely relate to IoT technology and in the verification and testing techniques. These can be related to the IoT systems. Bakar and Selamat \cite{bakar2018agent} conducted a systematic literature review and mapping study of the verification techniques used in the agent systems area. The study covers various types of verification techniques used during the design, development and runtime phases of the project. Methods analyzed in this study are applicable to IoT systems in general; however, the study is, by its design, limited to the agent systems.

Model-based Testing (MBT) represents a significant stream in system verification techniques. MBT is subject of another recent systematic literature review by Khan \textit{et al.} \cite{khan2018empirical}. This study focuses on the empirical verification of MBT techniques and concludes that the overall quality of reporting details can be improved in the significant part of the analyzed studies. Despite the fact, that the discussed MBT techniques are applicable in the IoT domain, this study is not specifically IoT-focused and the majority of the analyzed papers are dedicated to standard software systems, or describe general MBT techniques. 

Another related area is the early verification of SUT models created from the business requirements in a design phase of the project. A systematic literature review by Amjad \textit{et al.} \cite{amjad2018event} discusses these verification methods in event-driven process chain, which can be used in the design of IoT systems. Despite this applicability, the study is not directly focused on the IoT domain, so it does not discuss IoT--related specifics of the verification methods.

As security and privacy are two of the most critical aspects discussed in the current IoT systems \cite{Marinissen2016,Foidl2016,Kiruthika2015,trnka2018survey}, authentication schemes and methods are assessed as part of the QA process. A recent systematic literature review by Velasquez \textit{et al.} \cite{velasquez2017authentication} analyzes and categorizes the current authentication methods applicable to IoT systems. The topic is closely relevant to the IoT security and privacy issues. However, the study is focused on the general analysis of authentication schemes and these schemes are not discussed in the IoT context. Also, testing and quality aspects are not included in the scope of the study.

Regarding the privacy aspect, anonymous communication systems are raising the user's interest in the recent period. A systematic literature review by Nia and Martinez \cite{nia2018systematic} gives an overall picture of the state--of--the--art in this area and discusses its future research directions. In these directions, the quality of service is discussed; however, the work does not directly focus on other quality aspects. Moreover, the study is designed to analyze the general anonymous communication systems, which can be considered as a domain having an intersect with IoT systems.

In contrast to those mentioned systematic mapping or survey studies, this systematic mapping study looks directly into the QA aspect of IoT. 
As discussed previously, White \textit{et al.} \cite{WHITE2017186} addressed the quality of IoT from the service perspective that the application provides. The study considered the quality models that the ISO/IEC 25010 provides. The study tried to address three limited and straightforward research questions. In contrast to this approach, we are looking at the quality approach of the IoT from the system perspective. Here, we recognized those quality aspects considered when testing and evaluating IoT systems not necessarily from the service perspective. In line with this approach and with other research, the study tries to identify the different QA aspects addressed in the literature. In addition to the aims mentioned above, the study also seeks to answer many research questions regarding the QA aspects of IoT that have not been addressed so far.

\section{Method}

\label{Method}

\begin{figure}[h]
\centering \includegraphics[width=3.1in]{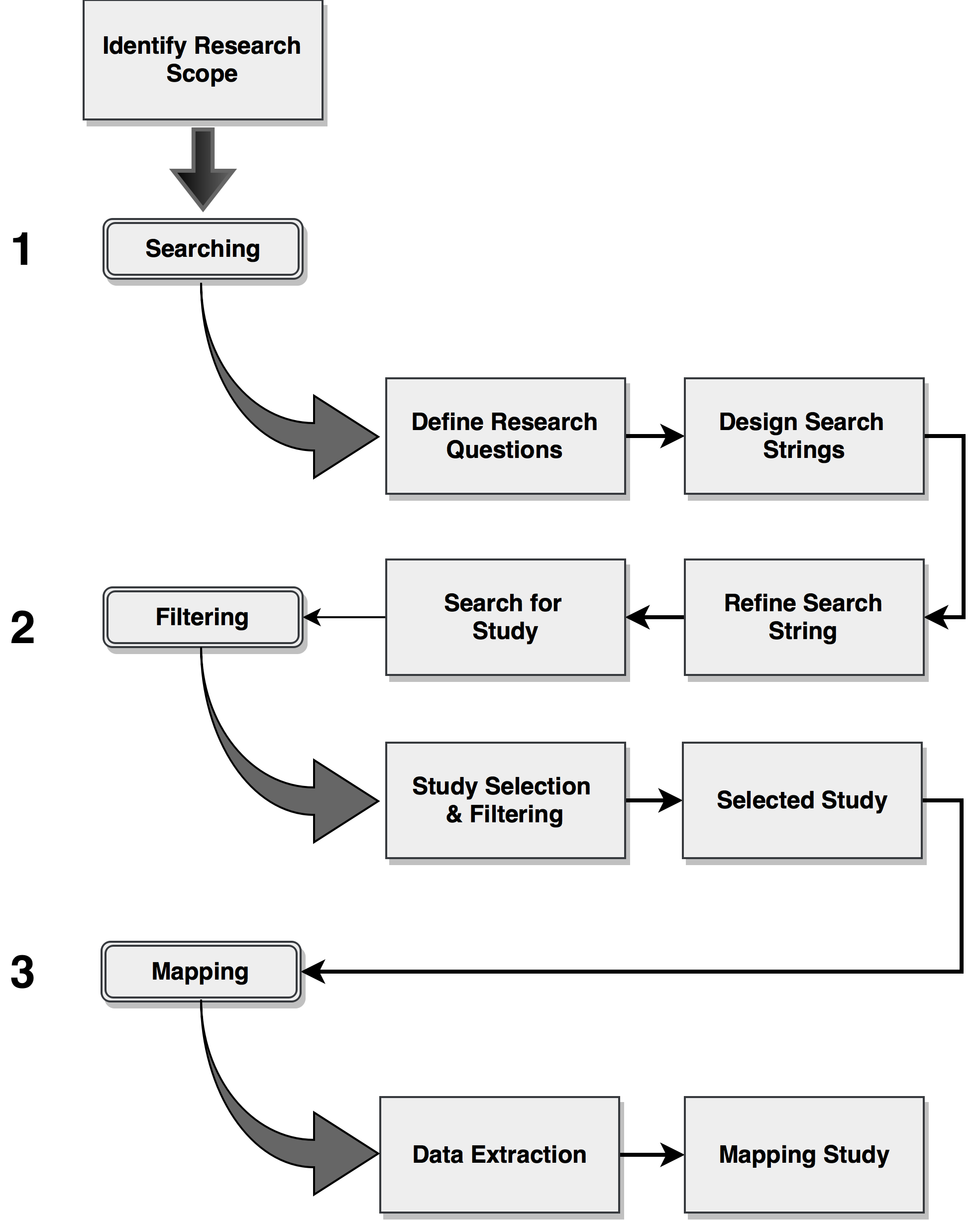} \caption{The systematic mapping detail steps\label{fig:The-systematic-mapping-figure}}
\vspace{-2em}
\end{figure}

This section explains the method used in this mapping study. The applied method based on guidelines provided by Petersen \textit{et al.} \cite{Petersen2015,Petersen2008}. The study first starts by identifying the scope. This research is only considering published papers that are related to the quality aspects of IoT. As can be seen in Figure \ref{fig:The-systematic-mapping-figure}, the study is composed of three main phases, each one of them has different stages. These phases are as follows:

\begin{enumerate}
\item \textbf{Searching Phase}: The research questions that determine the focus of the study are defined in this phase. Based on these research questions, the search string is designed. The search string has undergone different refinement process to identify and return the right papers. 
\item \textbf{Filtering Phase}: Here, the relevant papers are selected, and their quality is assessed. The papers are excluded from the primarily selected papers based on the title, abstract, full--text reading and quality assessments.
\item \textbf{Mapping Phase}: The relevant data answering the research questions are extracted from the primarily selected papers in this phase. The extracted data from the selected papers are classified to visualize the outcome. Here, tables and illustrations are used. Threats to validity are also analyzed and presented in this phase with the aim to demonstrate the possible limitations of the study.
\end{enumerate}

The following subsections illustrate each of the phases mentioned above in further detail. From Figure \ref{fig:The-systematic-mapping-figure}, we can clearly identify the input and output of each stage in the methodology. Some of the stages are included together in one subsection for the explanation. For instance, before identifying the research questions, it is necessary to know the research scope first. Hence, we have included both steps in one subsection. 

Within the answer of the RQs, (especially RQ3,4 and 6), we have also addressed the key research papers that illustrate the answers. In fact, the purpose of the paper is not to review each published paper in this direction; we are rather aiming to identify the future directions and the main aspects of research in addition to mapping them. This is considered a primary and essential study scope here.

\subsection{Research Questions\label{subsec:Research-Questions}}

For this study, multiple research questions have been raised, and different evidence examined. The study attempted to answer different questions that lead to a better understanding of the research notions and the future directions.  Particularly, the study tries to answer the following questions:
\begin{itemize}
\item \textbf{RQ 1:} What was the evolution in the number of published studies over the last decade for QA of IoT?
\item \textbf{RQ 2:} Which individuals and countries are active in conducting QA research for IoT?
\item \textbf{RQ 3:} Which aspects of the IoT quality have been dealt with in previous research?
\item \textbf{RQ 4:} Which principal testing techniques or concepts have been previously researched in the context of IoT?

\item \textbf{RQ 5:} Which specific application domains of IoT systems have been investigated from a quality viewpoint?
\item \textbf{RQ 6:} What are the current limitations and challenges in QA for IoT?

\end{itemize}

\subsection{Search Strategy\label{subsec:Search-Strategy}}

As previously mentioned, the scope of this study is the research papers related to the quality aspects of IoT. To define the right keywords for this scope this work considers the formal approach that establishes the Population, Intervention, Comparison, and Outcomes (PICO) criteria \cite{Kitchenham2010}. The Population represents the field discipline. The Inventions points to the methodologies and approaches to address the given issue. The Comparison considers available methods used for comparison. Finally, the Outcomes are the results for the readers and practitioners and help them with reaching the information.

Keywords can be categorized into three groups based on the research questions and the PICO. The primary group scopes the search for the QA in the IoT, such as ``quality assurance in the Internet of Things.'' Broadening the search the secondary group consists of terms and strings related to the testing domain. The last group is the application of the QA in the IoT, such as ``overload avoidance''. All these strings are combined to form strings with variations while combining search terms with logical AND and OR operators.

To form the final search string multiple preliminary attempts are made. The point of these attempts is to separate out QA from other domains not related to IoT. The intention is to narrow up the scope of the results. We have evaluated the search strings based on the quality of the returned results, which was assessed by the number of papers related to the scope of the study. For this quality assessment process of the search string, we have randomly selected 50 papers to search on IEEExplore and ScienceDirect and evaluated the search strings. The final test string was selected based on the results matching the expected scope. Also, we have assessed the quality of the search string based on the missing papers in the pilot set. Table \ref{tab:Search-string-tries} shows the result of each try of the search string including the results returned the number of missing papers for five attempts using different search strings. 

\begin{table*}
\caption{\label{tab:Search-string-tries}Search string tries on the indexing
data bases}
\centering{}%
\begin{tabular}{|l|>{\raggedright}p{8.5cm}|>{\raggedright}p{1.6cm}|>{\raggedright}p{1.6cm}|}
\hline 
Keyword  & Search Strings  & \# Results  & \# Missing Studies \tabularnewline
\hline 
Set \#1  & (\char`\"{}Internet of Things\char`\"{} OR \char`\"{}IoT\char`\"{})
AND (\char`\"{}Testing\char`\"{} OR \char`\"{}Quality\char`\"{}) & 1,513 & 22\tabularnewline
\hline 
Set \#2  & (\char`\"{}Internet of Things\char`\"{} OR \char`\"{}IoT\char`\"{})
AND (\char`\"{}Testing\char`\"{} OR \char`\"{}Quality\char`\"{} OR
\char`\"{}Security\char`\"{}) & 3,132 & 18\tabularnewline
\hline 
Set \#3  & (\char`\"{}Internet of Things\char`\"{} OR \char`\"{}IoT\char`\"{})
AND (\char`\"{}Testing\char`\"{} OR \char`\"{}Quality\char`\"{} OR
\char`\"{}QA\char`\"{} OR \char`\"{}Quality Assurance\char`\"{} OR
\char`\"{}Reliability\char`\"{} OR \char`\"{}Verification\char`\"{}
OR \char`\"{}Validation\char`\"{} OR \char`\"{}Testware\char`\"{}
OR \char`\"{}Testing Data\char`\"{} OR \char`\"{}Testbed\char`\"{}
OR \char`\"{}Performance\char`\"{} OR \char`\"{}Security\char`\"{}) & 3,1746 & 13\tabularnewline
\hline 
Set \#4 & (\char`\"{}Internet of Things\char`\"{} OR \char`\"{}IoT\char`\"{})
AND (\char`\"{}Test\char`\"{} OR \char`\"{}Tests\char`\"{} OR \char`\"{}Testing\char`\"{}
OR \char`\"{}Quality\char`\"{} OR \char`\"{}QA\char`\"{} OR \char`\"{}Quality
Assurance\char`\"{} OR \char`\"{}Reliability\char`\"{} OR \char`\"{}Verification\char`\"{}
OR \char`\"{}Validation\char`\"{} OR \char`\"{}Testware\char`\"{}
OR \char`\"{}Testing Data\char`\"{} OR \char`\"{}Testbed\char`\"{}
OR \char`\"{}Performance\char`\"{} OR \char`\"{}Security\char`\"{}
OR \char`\"{}Privacy\char`\"{})  & 3,560 & 7\tabularnewline
\hline 
Set \#5  & (\char`\"{}Internet of Things\char`\"{} OR \char`\"{}IoT\char`\"{})
AND (\char`\"{}Testing\char`\"{} OR \char`\"{}Quality\char`\"{} OR
\char`\"{}QA\char`\"{} OR \char`\"{}Quality Assurance\char`\"{} OR
\char`\"{}Reliability\char`\"{} OR \char`\"{}Verification\char`\"{}
OR \char`\"{}Validation\char`\"{} OR \char`\"{}Testware\char`\"{}
OR \char`\"{}Testing Data\char`\"{} OR \char`\"{}Testbed\char`\"{}
OR \char`\"{}Performance\char`\"{} OR \char`\"{}Security\char`\"{}
OR \char`\"{}Privacy\char`\"{} OR \char`\"{}Benchmark\char`\"{}) & 4,698 & 0\tabularnewline
\hline 
\end{tabular}
\end{table*}

As can be seen from Table \ref{tab:Search-string-tries}, the first four sets of strings were excluded since they produce many irrelevant results. The reason is the use of general terms. Moreover, some terms such as ``quality assurance'' is employed in different ways based on the context. Finally, different terms were used, (quality assurance OR quality measure OR quality evaluation), which lead to revealing more papers. Next observation is that these terms were used with different conditions than quality assurance. As a result, alternative terms were used as well. Ensuring the coverage of RQs and the search breadth, additional aggregation of terms was performed, such as ``strategy, technique, method, approach, and tool''. Besides, we noticed that some of our pilot set of papers were missing in the first four sets of strings as can be seen in the table. For these reasons, the fifth set of the search string is used for this research. 

The database selection was based on the guidelines and suggestions provided by \cite{Dyba2007,Petersen2008,Petersen2015}. This led us to the IEEE Xplore, ACM Digital Library, SpringerLink, and ScienceDirect databases. During the searching, indexing, and sorting of a vast number of references, multiple duplicate references came to place due to the tight variations in the reference indexing in different databases. To avoid duplication, references manager software EndNote X7 was used. To further improve accuracy, Mendele v1.16 reference manager software is used to refine the results.

The study maps research for the past years starting from 2009 to 2017. Since then there has been increasing research trend. It should be mentioned that this study started in 2018. Thus, papers published in 2018 are excluded. Table \ref{tab:Numbers-of-published-papers-per-database} summarizes the number of research papers published in the mentioned period for each considered database.

\begin{table}[b]
\vspace{-1.5em}
\centering{}\caption{\label{tab:Numbers-of-published-papers-per-database}Numbers of published
research}
\begin{tabular}{|l|l|}
\hline 
Database  & Search results \tabularnewline
\hline 
IEEE Xplore  & 9,175 \tabularnewline
\hline 
ScienceDirect  & 6,030 \tabularnewline
\hline 
ACM DL  & 1,288 \tabularnewline
\hline 
SpringerLink  & 13,471 \tabularnewline
\hline 
Total  & 29,964 \tabularnewline
\hline 
\end{tabular} 
\end{table}

\subsection{Paper Selection Criteria and Quality Assurance\label{subsec:Article-Selection-Criteria}}

The papers found by search strings were selected or excluded based on the title, abstract and full--text reading. During this selection, the quality of the papers was also taken into account. To increase the reliability of this process and to reduce possible threats of subjective selection, this process was conducted by the first and second authors and reviewed by the other authors of this study. Some papers could have been selected or excluded based on the title and abstract. Nevertheless, full--text reading was required for some papers to determine if the paper is relevant for the selection. In this mapping study, the papers fulfilling the following criteria were selected:

\begin{enumerate}
\item Papers discussing quality assurance aspects of IoT solutions in general
terms. 
\item Papers focusing on a particular aspect of IoT quality.
\item Industrial case studies, where quality aspects are discussed. 
\item Methodologies for IoT quality assurance. 
\item Papers with full text available in the selected databases. 
\item Papers published with full text online from 2009 to 2017. 
\end{enumerate}

By quality aspects and quality assurance aspects we consider any aspect related to IoT solution reliability, correctness, and durability, together with security and privacy aspects. Following the guidelines provided by Petersen \textit{et al.}  \cite{Petersen2008,Petersen2015}, papers meeting the following criteria were excluded:

\begin{enumerate}
\item Papers not exactly related to the scope of this paper, i.e., papers describing aspects of IoT other than quality. 
\item Papers not presented in the English language. 
\item Papers without full text available in the selected databases. 
\item Books and gray literature. 
\item Papers from non--peer reviewed sources. 
\end{enumerate}

\begin{figure}
\centering
\includegraphics[scale=0.45]{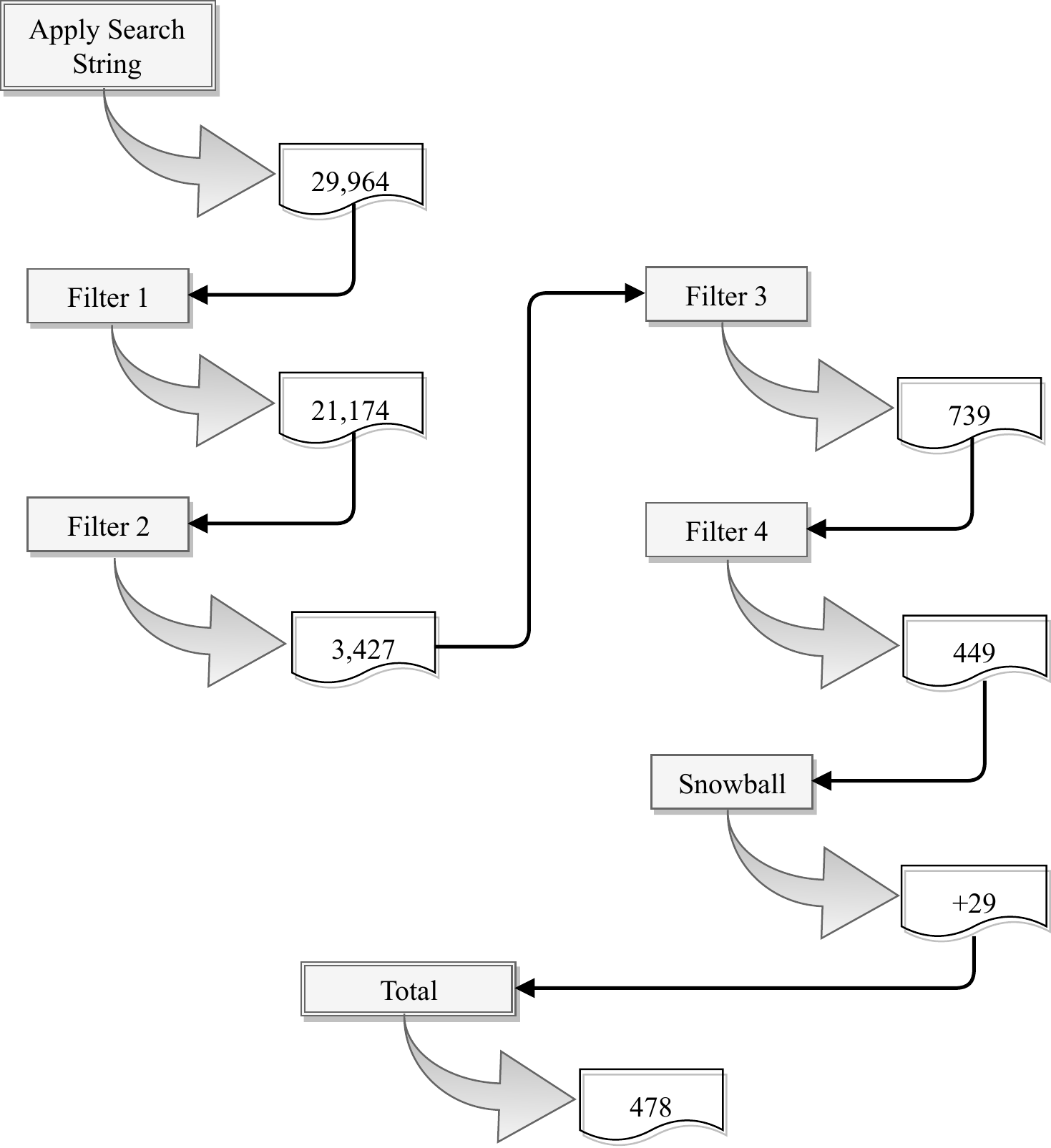}

\noindent \caption{\label{fig:Filtering-stages}Filtering stages of the selected papers and the number of papers in each stage}
\vspace{-1em}
\end{figure}

Following the criteria mentioned above for selection and exclusion, we have considered many useful papers for our study. In fact, some of those papers were duplicated in the selected databases. For example, papers that appeared in ScienceDirect were also listed in ACM. Figure \ref{fig:Filtering-stages} shows the stages we followed to get the final set of the studied papers. 

Following the robust methodologies of other mapping studies \cite{daMotaSilveiraNeto,Alves2016}, the selection process of the final papers underwent different filtering stages. The selection process started by searching for the papers in the IEEE Xplore, ScienceDirect, ACM Digital Library, and SpringerLink databases with the chosen search string in Section \ref{subsec:Search-Strategy}. The outcome of this stage was 29,964 papers. The broad range of articles considering IoT in different parts of the text resulted in this enormous number of papers. For example, we found many papers mentioning the usefulness of specific solutions for IoT applications in the conclusion section. However, none of these papers were related to the quality aspects of IoT. After retrieving the full list of papers, during the ``Filter 1'' stage, we removed the duplicates --- the papers shared among the databases. 

During ``Filter 2'' stage, we followed the inclusion and exclusion criteria described earlier. Here, the decision depended on the paper's title, reading abstract and if necessary reading the introduction section. 

In ``Filter 3'' stage, we excluded other papers based on the full--text reading. Here, we excluded those papers without a particular method and experiments. We noticed that a huge number of papers was published only for illustrating possibilities of using IoT concepts in different applications. There is also a huge number of papers describing frameworks for IoT applications without performing any verification and validation experiment. For example, we noticed many papers describing security and privacy frameworks for IoT solutions. Also, we have also excluded pure review papers. 

The final filtering stage ``Filter 4'' was related to the quality of the published papers. We noticed many papers with low--quality contents that have been coming from unreliable and non--reputable conferences. Another set of low--quality papers were coming from non--peer reviewed journals. For better reliability of the results, each author conducted a snowballing search for other possible missing relevant papers. Here, we have added 29 more papers. We end up with 478 papers that need to study for answering our established research questions. The online mapping resources and the full list of papers with all the details can be found in an online Spreadsheet\footnote{http://bit.ly/2NDgpZk}. 

\subsection{Data Extraction and analysis\label{subsec:Data-Extraction-and}}

In this stage of the study, we extracted the data from the selected set of papers. This stage aimed to map and classify the papers to enable us handling the RQs addressed in Section \ref{subsec:Research-Questions}. For better organization and systematic flow of the work, we have created a spreadsheet template (See Table \ref{tab:Data-Extraction-Template}) with all the relevant information about each paper. The sheet is an extended sheet with more details that have been originally presented by \cite{Petersen2015,Petersen2008,deMagalhaes2015}. For each paper, we filled the sheet with paper ID, publication title, publication year, authors\textquoteright{} names and countries, venue, and the area of research. To extract and analyze the information from the sheet, we took two directions, manual and dynamic extraction. The first and fourth authors did the manual extraction and reviewed by the other authors. For double check and reliability, we have used mining and automatic text analyzers also for verification. 

\begin{table*}
\vspace{-1em}
\caption{\label{tab:Data-Extraction-Template}Data Extraction Template}

\centering
\centering{}%
\begin{tabular}{|l|l|}
\hline 
Data item & Value\tabularnewline
\hline 
\hline 
Study ID & Integer\tabularnewline
\hline 
Paper Title & Name of the paper\tabularnewline
\hline 
Author Name & Name of author(s)\tabularnewline
\hline 
Year of Publication & Calendar year\tabularnewline
\hline 
Venue & Name of publication venue\tabularnewline
\hline 
Country & Name of the country for each participated author\tabularnewline
\hline 
Area of research & Knowledge area of research\tabularnewline
\hline 
Research topic & Main topic or theme addressed by the study\tabularnewline
\hline 
Research problem & Research problem addressed by the study \tabularnewline
\hline 
Proposal  & Proposed solution to the problem\tabularnewline
\hline 
Contribution  & Main contribution of the paper \tabularnewline
\hline 
Evaluation process & Which benchmark adopted for evaluation?\tabularnewline
\hline 
Case study & Which case study used?\tabularnewline
\hline 
\end{tabular}
\end{table*}

\section{Results\label{sec:Results}}

By extracting the information from the selected papers, we can answer the RQs raised earlier. We have addressed the answer for the RQs individually. The following subsections illustrate the results of the study and the answer for each RQ. For abstraction, we used a short title for the sections that have been extracted from the main RQs. 

\subsection{Frequency of publication (RQ1)}

\begin{figure}
\begin{centering}
\includegraphics[scale=0.75]{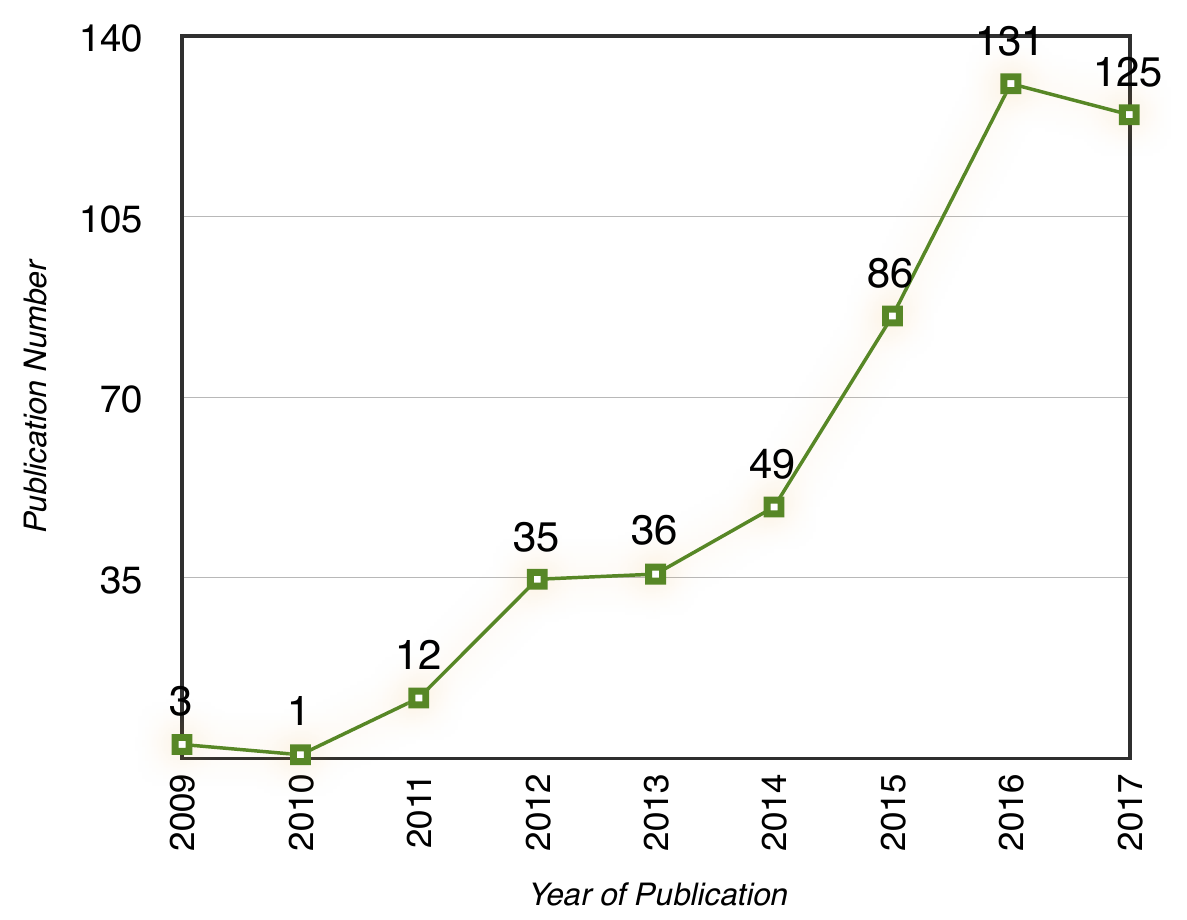}
\par\end{centering}
\caption{\label{fig:Publication-per-year}Publication per year}
\vspace{-1em}
\end{figure}

By analyzing the selected papers from 2009 to 2017, we can answer the RQ about the frequency of publication and also the evolution in the publication number. Figure \ref{fig:Publication-per-year} illustrates this evolution by showing the number of published papers per year. As we considered 478 papers for this study, we can observe that the average publication number per year is almost 53 papers starting from 2009 in which the first set of papers published.

\begin{figure}
\includegraphics[scale=0.7]{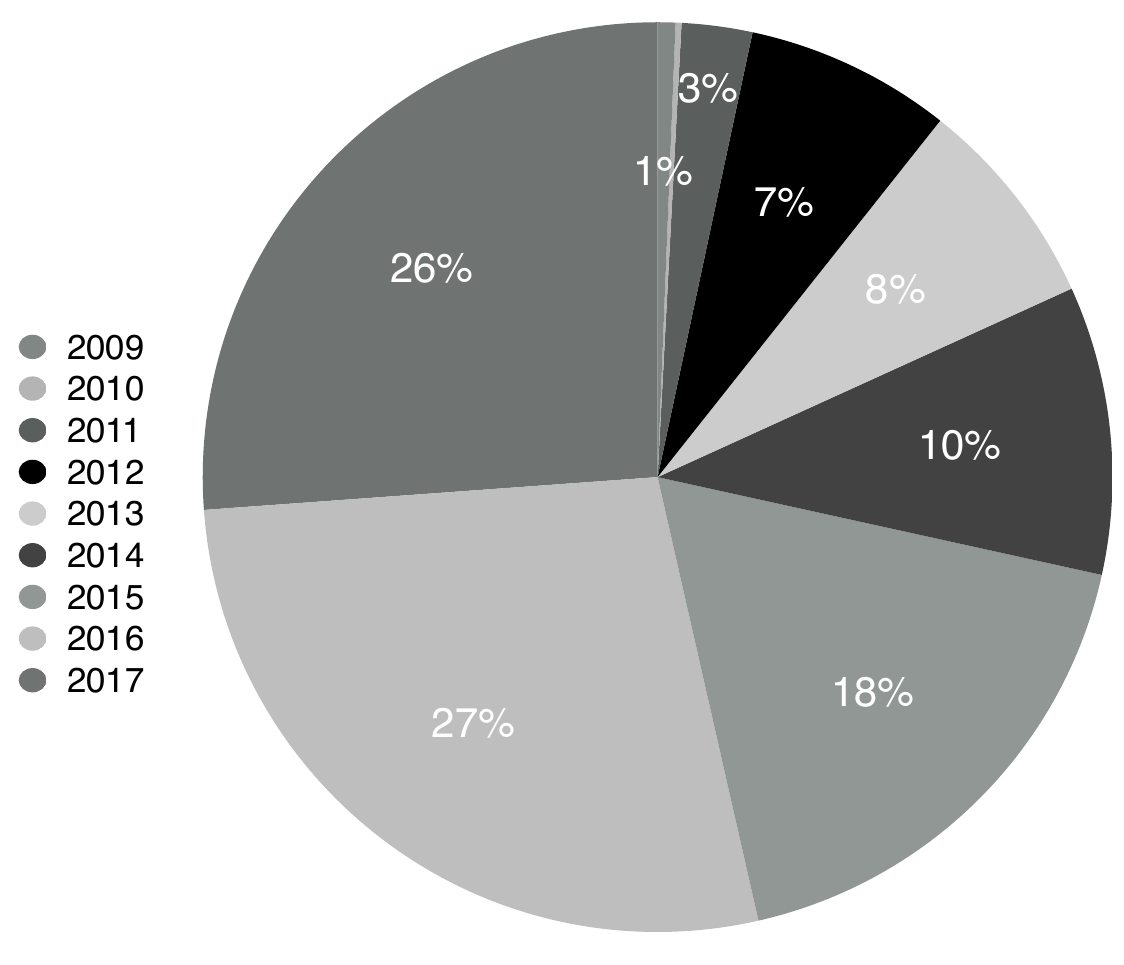}
\centering{}\caption{\label{fig:Publication-ratio-per-year}Publication ratio per year}
\vspace{-1em}
\end{figure}

Although the name IoT and the research related to it started in 1999, we can see that researchers have begun to publish papers about its quality aspects since 2009 actively. Figure \ref{fig:Publication-ratio-per-year} shows publication ration per year. In both Figures \ref{fig:Publication-per-year} and \ref{fig:Publication-ratio-per-year}, we can see the number of published papers such that in 2016 there were 131 papers published which is almost  27\% of the total published papers and in 2017; there were 125 papers published, which is almost 26\% of total published paper volume. From the figures, it is clear that the interest of IoT quality aspects publication is increased in the research community after 2010. An important potential reason behind this increase is the emerge of many new IoT solutions for daily life applications and the influence of these applications in our daily lives. 

\begin{figure*}
\vspace{-1em}
\begin{centering}
\includegraphics[scale=0.9]{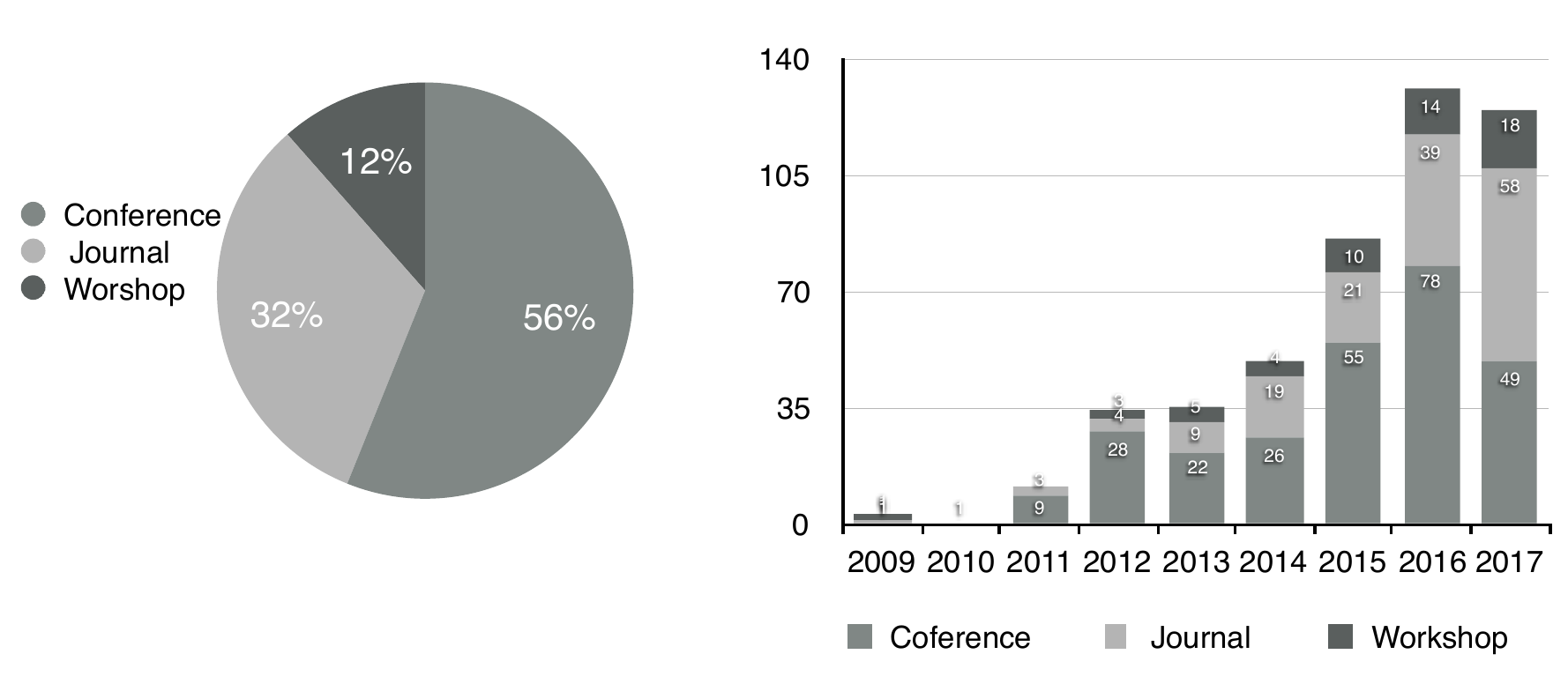}
\par\end{centering}
\vspace{-1em}
\caption{\label{fig:Publication-number-and-ratio}Publication number and ratio
categorized by publication type}
\vspace{-1em}
\end{figure*}

It is also important to analyze the type of publication venues and also the ration of publication for each venue. This information can be extracted clearly from Figure \ref{fig:Publication-number-and-ratio}. We can see that the majority of the papers (approximately 56\%) published in conferences, while approximately 32\% and 12\% published in journals and workshops. Figure \ref{fig:Publication-number-and-ratio} also gives the analysis of publication type based on the venue per year. We note that the majority of the conference, journal, workshop papers was published in 2015, 2016 and 2017. 

\begin{figure*}
\vspace{-1em}
\begin{centering}
\includegraphics[scale=0.9]{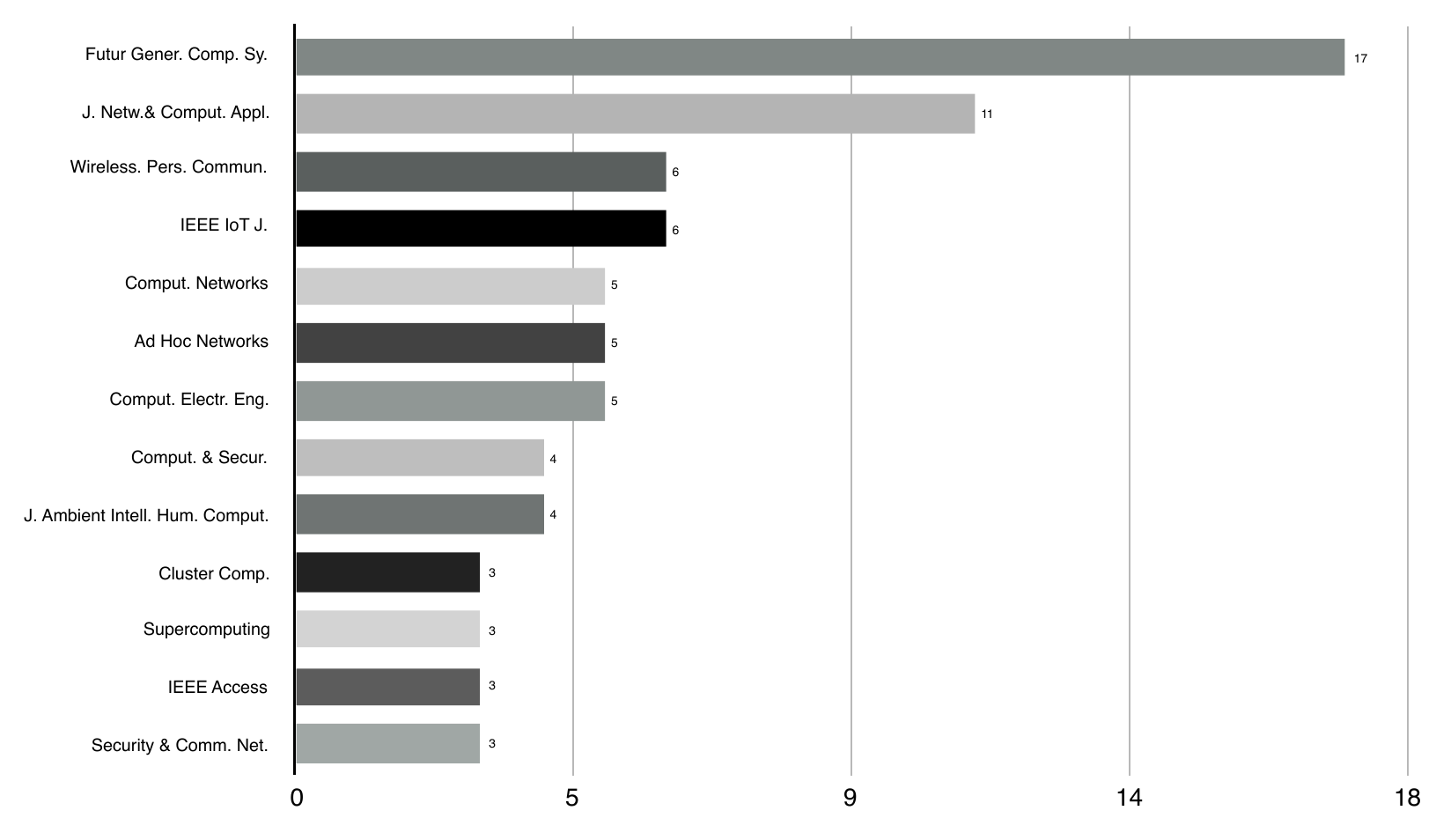}
\par\end{centering}
\vspace{-1em}
\centering{}\caption{\label{fig:Amount-of-Published-vs-JournalsNames}Amount of Published
Articles vs. Journal Name}
\end{figure*}

Another important observation from this analysis is the favorite and frequent peer--reviewed journals, conference, and workshops in which the papers have frequently been published. Figure \ref{fig:Amount-of-Published-vs-JournalsNames}, shows those favorite journals with the number of published papers for each journal. We have used Thomson Reuters Science Citation Index\footnote{https://apps.webofknowledge.com} for the journal abbreviation. The full name of journals can be found in Appendix A. Additionally, Figure \ref{fig:Amount-of-Published-vs-Conference-venues} shows those active and popular conferences in which the related papers published. Note, we used conference abbreviation for the name and the full names could be found in Appendix B. It should be mention here that we considered a journal or conference to be popular when more than two related paper published in it. 

\begin{figure*}
\begin{centering}
\includegraphics[scale=0.8]{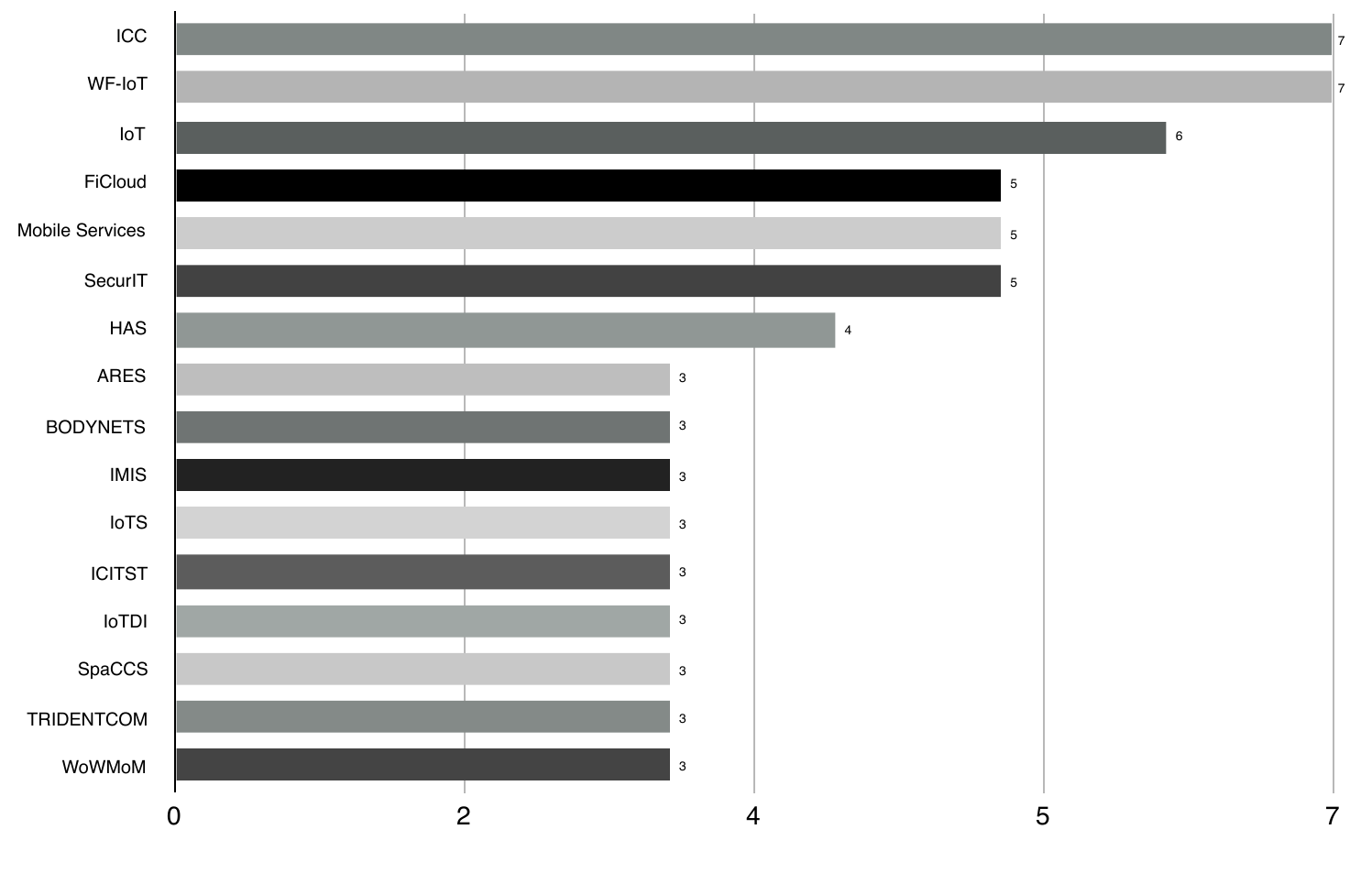}
\par\end{centering}
\vspace{-1em}
\centering{}\caption{\label{fig:Amount-of-Published-vs-Conference-venues}Amount of Published
Articles vs. Conference Venues}
\vspace{-1em}
\end{figure*}

We can observe from Figures \ref{fig:Amount-of-Published-vs-JournalsNames} and \ref{fig:Amount-of-Published-vs-Conference-venues} the targeted journals and conference by the author to publish papers related to the quality aspect of IoT. We can see that ``Future Generation Computer Systems journal'' with 17 published papers is the most popular journal in this field. Also, ``Journal of Network and Computer Applications'', ``Wireless Personal Communications'', and `IEEE Internet of Things Journal'' with 11, 6, and 6 published papers are the second and third popular journals in the field. The other set of the most active journal set is ``Computer Networks'', ``Ad Hoc Networks'', and ``Computer and Electrical Engineering'' with five published papers for each one of them. The journals with more than two papers published in this field form more than 48\% (75/154) of the published journal papers and more than 15\% (75/478) of the whole published papers. 

Looking at the conference venues, we can see that ``International Conference on Communications (ICC)'', ``IEEE World Forum on Internet of Things (WF--IoT)'', and ``International Conference on the Internet of Things (IoT)'' are the most three active and targeted conferences by authors with 7, 7, and 6 published papers. These three conferences form more than 7.4\% (20/269) of the conferences publication. Also, we noticed that those conferences with more than two published papers form more than 13.8\% (66/478) of the publication. However, there are many papers related to the quality aspect of IoT published in individual conferences.

\subsection{Active Individuals, Organizations and Countries (RQ2)}

\begin{figure*}
\begin{centering}
\includegraphics[scale=0.8]{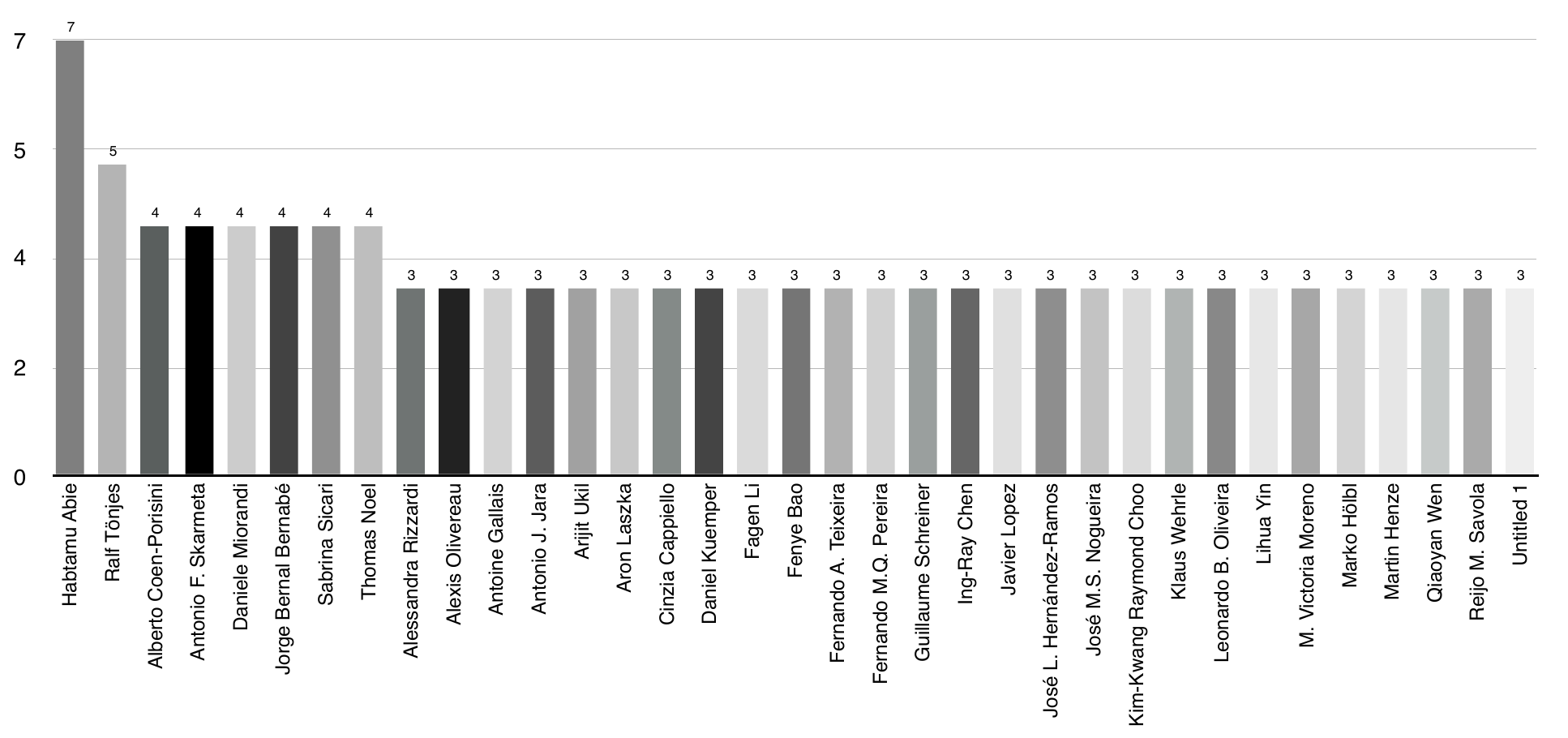}
\vspace{-2em}
\par\end{centering}
\caption{\label{fig:Active-Researchers}Active Researchers}
\vspace{-1em}
\end{figure*}

Within the answer of this RQ, we try to know those active researchers who published research papers related to any aspect of quality for IoT. By analyzing the frequent author names for the chosen population of papers, we found those active researchers. Here, we can define active researchers as those researchers (author/co--author) which are participating in more than one published paper. Figure \ref{fig:Active-Researchers} shows the ranking of those active researchers showing their full names. 

As can be seen from Figure \ref{fig:Active-Researchers}, ``Habtamu Abie'' from Norwegian Computing Center and ``Ralf T{\" o}njes'' from ``University of Applied Sciences Osnabr{\" u}ck, Germany'' are the most active researcher by publishing seven and five papers. Also, as we can see from the figure, six researchers have published and participated in four papers. Based on our analysis, we should mention here also that, 27 individual authors are participating in more than two published papers, 111 individual authors are participating in more than one paper, and 1416 individual authors are participating in one or more published papers related to different quality aspects of IoT.

\begin{figure*}
\begin{centering}
\includegraphics[scale=1]{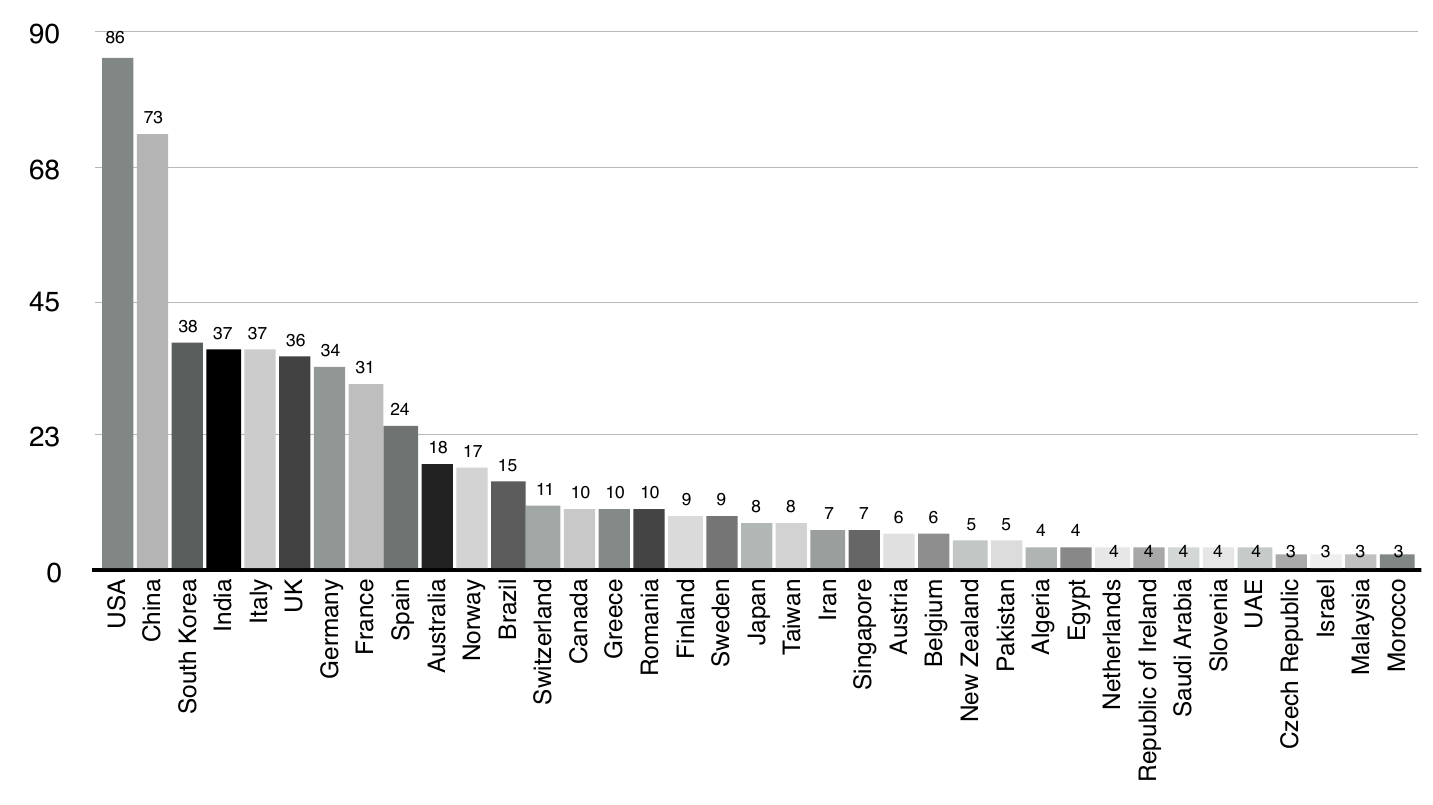}
\par\end{centering}
\vspace{-2em}
\caption{\label{fig:Active-countries}Active countries}
\vspace{-1em}
\end{figure*}

Another valuable information can be obtained from the analysis of the extracted information from the papers is the list of participating countries. Figure \ref{fig:Active-countries} shows the result of this analysis. We took the participation of the countries based on the author's organization affiliation on each published paper. Each country is counted one time per paper. For example, if three authors from the USA participated in a paper, we count the USA for one time. Based on the results shown in Figure \ref{fig:Active-countries}, we noticed that USA, China, and South Korea are the most active three countries in research publication related to different quality aspects of IoT by publishing 86, 73, and 38 papers respectively. These three countries together form more than 41\% (197/478) of the whole publication. However, we also noticed other active countries that can be competitive in term of research numbers since they are so close to each other. These countries are India (37 papers), Italy (37 papers), United Kingdom (36 Papers), Germany (34 papers) and France (31 papers).

The output of this research question can reveal important pieces of information. The number of participated researchers in publication regarding different quality aspects of IoT shows a promising and important research direction. However, we noticed that there is no such active organization or research group that focusing on IoT quality. 

\subsection{Aspects of IoT and Topics Addressed (RQ3)}

\begin{figure}
\includegraphics[scale=0.58]{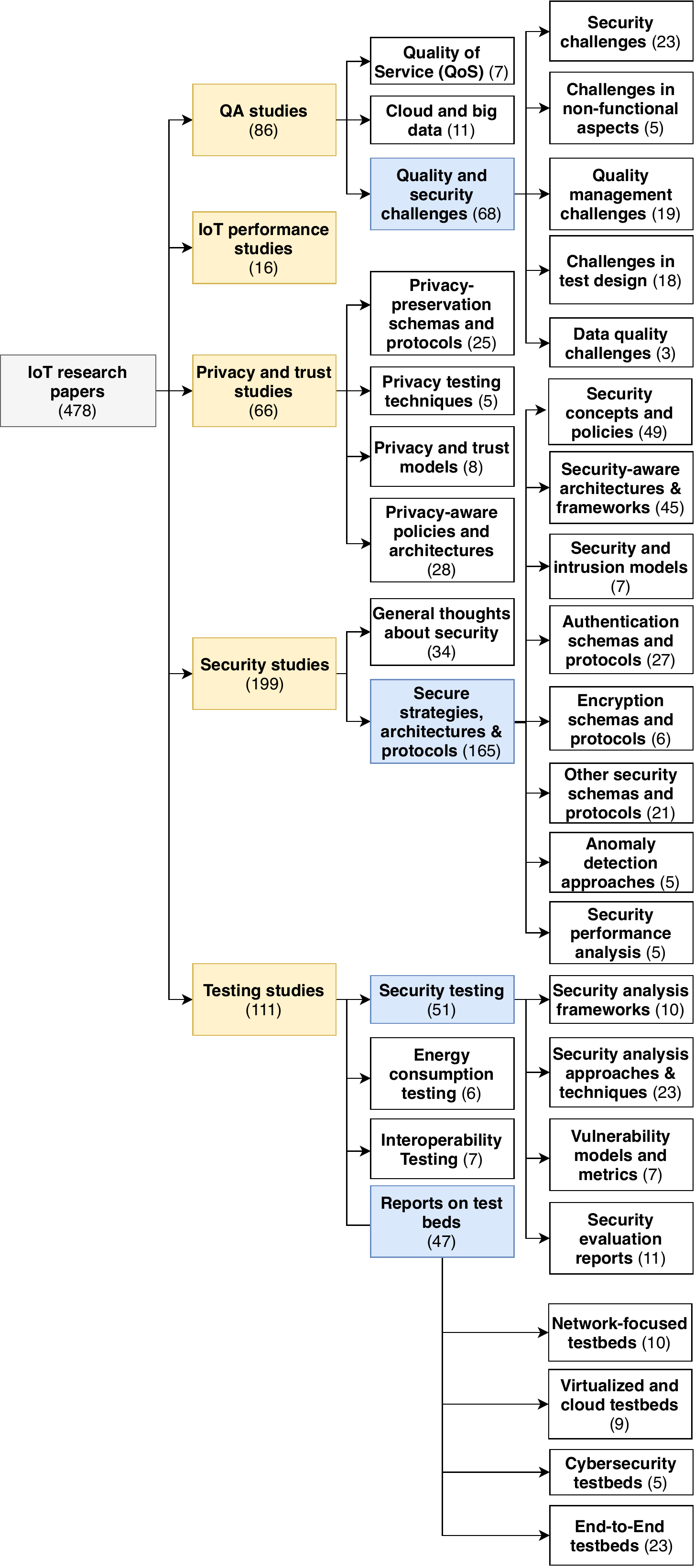}
\vspace{-0.5em}
\caption{\label{fig:Top-view-classification}Top view classification for the IoT quality--related topics}
\vspace{-1.3em}
\end{figure}

It is impossible and unrealistic to cover all the detail issues, aspects, and topics related to the quality of IoT in one paper. However, we analyzed those selected papers and came out with a high--level classification that provides an essential overview of key aspects of quality for IoT. It should be mentioned here that these aspects are the aspects that were the focus of published research papers; however, many other aspects gain less attention in the published paper. Figure \ref{fig:Top-view-classification} shows a top view classification tree for the topics dealt with in the analyzed papers including the number of published papers for each topic.

\subsubsection{QA studies}

QA studies are going side by side with testing studies. Generally, QA studies work on defining frameworks to assure the quality of selected aspects of IoT systems or discuss quality--related issues in general.

In this context, several studies that focusing on quality of \textit{Cloud and big data} concepts can be identified (11 papers). This is a logical situation, as the cloud computing and big data concept closely related to a number of contemporary large IoT systems. On the network level, \textit{Quality of Service (QoS)} is discussed by seven papers. The majority of papers related to the QA studies category discusses various \textit{Quality and Security challenges}, namely 68 papers in total. In this category, several subcategories can be clearly identified. The most concerns are raised regarding the security of the IoT systems (category \textit{Quality and Security challenges}, 23 papers) and the topic is discussed from various aspects, including also user and data privacy (particular security and privacy studies are analyzed in subsections \ref{subsec:security_studies} and \ref{subsec:privacy_studies} of this chapter). Besides the security concerns, general \textit{Quality management challenges} is the next frequently discussed area, covered by 19 papers. This area includes test management, metrics as well as organizational and line management aspects of IoT quality. Preparation of effective and accurate test cases with adequate coverage is important for the overall efficiency of the testing process and challenges in this area are also discussed (category \textit{Challenges in test design}, 18 papers). Among the main challenges in the design of test cases, an excessive number of possible combinations to test and challenges in integration testing are reported. Besides the functional, integration and combinatorial testing, \textit{Challenges in non-functional aspects} of the IoT systems were discussed by five papers. Finally, \textit{Data quality challenges} were also subject of three analyzed studies.

As there are many applications in the IoT, it is impossible to determine general QA aspects. Here, methods of verification and validation are get involved in the research papers for specific purposes. The problem of underlying heterogeneous nature, platform variants, and type of IoT are some of those difficulties may arise when someone tries to define such a QA process. Though, few research studies attempted to follow this direction by introducing new methods for how to assure the quality of particular IoT service and platform.

Reetz \textit{et al.} \cite{Reetz2013} described an approach to test IoT services based on the code insertion methodology to address the interaction with the physical world. To test the applicability and efficiency of classical approaches, the study emulated the IoT resources from an implementation perspective. Assuring the physical world and the real implementation of the IoT services and ``Things'' like sensors is not an easy task since it is difficult all the factors that affect the actual implementation in the emulator. To overcome this variation between real implementation and the emulator, Gimenez \textit{et al.} \cite{Gimenez2013} designed and developed a simulation environment for IoT services to implement multiple types of sensors. Here, to test the quality from an implementation perspective, the study uses a standard sensor database called Sensor Observation Service (SOS).

Assuring the quality of service for cloud platforms when interacting with a large number of things is an important approach. This quality assurance process will be an essential differentiation metric to choose among the many cloud providers available nowadays. For this reason, Zheng \textit{et al.} \cite{Zheng2014} propose a quality model named CLOUDQUAL for cloud providers to assure the quality of service for IoT. The model can be used to measure, represent or compare the quality of different providers. The model takes six well-known quality dimensions which are reliability, availability, usability, security, elasticity, and responsiveness. While this study takes these six dimensions of quality, Ahmad \cite{AhmadIoT2014} has just concentrated on the reliability models in more detail. The study focuses on the prediction and estimation of hardware and software reliability by deriving a probabilistic estimation of overall system reliability. Karkouch \textit{et al.} \cite{Karkouch2016} going further with this approach to develop a Model-Driven Architecture-based approach for quality. Here, the data consumer will be given an opportunity to choose and illustrate his data quality requirement through models and a user-friendly graphical model editor. In the same way, Silva \textit{et al.} \cite{Silva:2013} created a tool to evaluate the dependability of IoT applications, which is defined in the study as reliability and availability.

Apart from the definition of these quality models, there are few studies towards the assurance of different IoT services. For example, \cite{Kim2016} worked on how to maximize the quality of information for IoT from a real-time scheduling perspective. Shi \textit{et al.} \cite{Shi2013} proposed an approach to design behavior patterns for the intelligent sensors in the IoT applications. The approach will help in the program implementation for these intelligent sensors which leads to improving the quality of service.

\subsubsection{IoT performance studies}

For those analyzed papers in this study (18 papers in total), we noticed that performance studies are mainly focused on the performance evaluation of different IoT streams. Those streams are distributed on performance evaluation of specific IoT platform, security protocols, network infrastructure, or particular medical or industrial application. In fact, we can see that there is no universal and unique method for IoT performance evaluation as there are many forms of it. This shows the difficulty of IoT performance evaluation since there is a possibility to develop an entirely different performance evaluation strategy for each IoT stream.

In addition to the quality assessment of a particular application of IoT, performance evaluation is important for many other reasons. For example, it is essential to know which protocol can provide the appropriate level of security for an IoT system. This cannot be done without evaluating and comparing different protocols. To this end, few studies have been conducted to provide the appropriate performance evaluation and comparison process for this reason. For example, Rubertis \textit{et al.}\cite{Rubertis2013} proposed a performance evaluation process to compare two well--known security protocols which are IPSec and Datagram Transport Layer Security (DTLS). The process can be applied to different protocols to provide the design the most appropriate and secure protocol for end--to--end IP communications for IoT. 

Performance evaluation is also used to assess the quality of IoT platforms (sometimes called framework). The IoT platform is a computational cloud middleware engine that manages the large number of data streams that coming from different sensors. Vandikas and Tsiatsis \cite{Vandikas2014} assessed these frameworks based on the throughput of the system as well as stability concerning robustness (i.e., dropped connections) and memory consumption. 

Performance evaluation processes have been designed specifically for individual IoT applications. For example, Yamada \textit{et al.} \cite{Yamada2016} designed a performance assessment method for IoT-based e-Learning test bed. The evaluation is focused mainly on the Optimized Link State Routing (OLSR) protocol. For the evaluation, the throughput, PDR, hop count, delay and jitter metrics are considered.

\subsubsection{Privacy and trust studies}
\label{subsec:privacy_studies}

In general, nowadays, Internet users need to use applications or devices that are connected to the Internet with a high level of privacy. This will enable the users to do daily activities with reliability and trust. In fact, IoT security is not far from this context, and it goes side by side with privacy. Users need to trust the IoT applications they use and to trust that the information generated by the IoT device is secure. In IoT applications, the number of connected devices could be large, and hence there is a need for a robust design for this issue at the system level and also for each device.

In the analyzed studies, we have identified several main streams of research. Papers in the \textit{Privacy--preservation schemas and protocols} category (25 papers) discusses various particular techniques, algorithms, schemas, and protocols to ensure data and user privacy in various IoT systems, covering IoT system users' personal data, as well as general business-domain data processed by the IoT system. Also, several particular testing techniques to assess the privacy level of IoT systems have been proposed (category \textit{Privacy testing techniques}, 5 papers). Besides that, several attempts to model the privacy and trust problem in an IoT system by a formal technique were published (category \textit{Privacy-aware policies and architectures}, 28 papers). These models can be further employed in the future development of privacy--aware architectures and various schemas an protocols. Finally, a considerable number of published studies discusses general processes, architectures, framework or policies to ensure privacy in an IoT system  (category \textit{Privacy-aware policies and architectures}, 28 papers). In contrast to \textit{Privacy--preservation schemas and protocols} category, the privacy issue is described in the more broad context of particular policy or system architecture, including various case studies. These studies also include general discussions of the privacy concepts in IoT systems.

Generally, privacy and trust studies in IoT are swirling around defining privacy policy and trust management of different applications. This includes defining various models and protection frameworks for privacy to increase the trust perceptions in the IoT. In this context, Sun \textit{et al.} \cite{Sun2014} propose a privacy protection policy to protect the security of personal information on IoT systems. The policy is mainly based on the homomorphism encryption algorithm. Samani \textit{et al.} \cite{SAMANI2015} proposed another policy for privacy but this time by modeling the IoT system first as a Cooperative Distributed Systems (CDS). Here, the CDS model has been analyzed then the privacy protection is recognized as a form of \textquotedblleft sensitive information\textquotedblright{} at the interactive level. In line with this modeling approach to privacy, Cao \textit{et al.}\cite{Cao2016} proposed another model for privacy but this time for data sharing in smart cities. The model covers the data abstraction and semantic, system architecture for data sharing, and strategies to enhance the transparency of data sharing without affecting the privacy. In fact, few research papers can be found in this direction, for example, \cite{Nitti2012,BenSaied2013,Hoepman2012,Schulz2013}.

Another set of research papers defined the location-based privacy for IoT systems. In this context, Liu \textit{et al.} \cite{JLiu2012} proposed a strategy for how to protect the user's location when there is a personalized service inside the IoT application. The strategy also contains a pseudonym policy and a model for location-based privacy by protecting the user's location information acquisition.

Not far from these important areas for privacy, there are plenty of research papers that define privacy for specific applications of IoT. For example, Ukil \textit{et al.} \cite{Ukil2015} defines the privacy issues within smart energy systems. Here, the study describes the uniqueness of privacy within smart energy systems from the smart meter (component of the smart energy management system) point of view. The meter could be a possible breaching activity for privacy when detecting in-house activities for example. Hence, the preservation of smart meter data would be essential. The study proposes a new scheme to minimize the privacy breaching risk in smart energy systems, called \textquoteleft Dynamic Privacy Analyzer\textquoteright{} scheme. In another study \cite{Gong2015}, the medical healthcare system is proposed for IoT to protect the privacy of patients' information in smart healthcare systems. Here, a private lightweight homomorphism algorithm is proposed that has been combined with an encryption algorithm.

\subsubsection{Security Studies}
\label{subsec:security_studies}

Security brings many concerns to IoT solutions e.g., \cite{Marinissen2016,Foidl2016,Kiruthika2015}. In total, we identified 199 studies in the whole body of the analyzed papers corresponds to the importance of this topic. Besides the general discussions about various security aspects and consequences in the IoT systems (category \textit{General thoughts about security}, 34 papers), 165 papers are discussing particular security strategies, architectures and protocols to ensure the desired level of security in the IoT systems. These studies span from reports on low--level security protocols to a description of general security--aware frameworks and security policies. Starting from the high--level conceptual descriptions, various \textit{Security concepts, and policies} are discussed by 49 papers, which are covering a variety of particular IoT domains, for instance, Wireless Sensor Networks, Smart Homes, Healthcare or Wearables and personal devices. More detailed security--aware solutions, architectures, and particular frameworks are described by the second largest subcategory of the papers in this area (category \textit{Security--aware architectures and frameworks}, 45 papers). Also here, these proposals are spanning among a number of various business domains and types of IoT systems, covering domains of Wireless Sensor Networks, Energy and Smart Grids, Healthcare, Smart Buildings, and Smart Homes as well as Smart Cars. Similarly, as in the Privacy area, models are also used to capture security issues and possible intrusion scenarios, seven analyzed papers were dedicated explicitly to the topic of \textit{Security and intrusion models}. A considerable number of papers is dedicated to various security--related schemas and protocols. These studies cover \textit{Authentication schemas and protocols} (27 papers), detailed presentation of \textit{Encryption schemas and protocols} (6 papers) and \textit{Other aspects of security schemas and protocols}, describing the problem from the broader perspective than only authentication or encryption mechanism (21 papers). Also, several \textit{Anomaly detection approaches} are discussed (5 papers). We used this subcategory to distinguish papers which present a runtime anomaly detection approaches and algorithms, not formulated as specific security testing techniques, which are further analyzed in Section \ref{subsec:testing_studies}. Besides, five papers were dedicated to \textit{Security performance analysis} area, also apart from specific security testing techniques discussed in Section \ref{subsec:testing_studies}.

In IoT security, unfortunately, the variety of devices and vendors of \textquotedblleft Things\textquotedblright{} make it hard to agree on how to implement security in devices. One cannot expect that existing security solutions that went through long evolution fitting to servers would also suit to IoT. IoT's have diametrical differences from centralized solutions involving client-server interaction. In this case, IoT is more similar to peer-to-peer (p2p) networks when it comes to security. A lot of research in p2p \cite{7284949} addresses security and defense mechanisms in the distributed environment. However, p2p networks do expect peers to possess reliable hardware, which is not the case for IoT. In p2p networks we usually think of a distributed overlay network of computers, however, in IoT, we interconnect \textquotedblleft small\textquotedblright{} things with limited processing power, which inhibits encryption and robust security measures. Vendors are pushing to reduce devices prices and rather focusing on sensors and data collection, which naturally leads to fewer efforts placed on security concerns. In IoT, there is no silver bullet to reduce security threats.

\begin{figure*}
\begin{centering}
\includegraphics[scale=0.36]{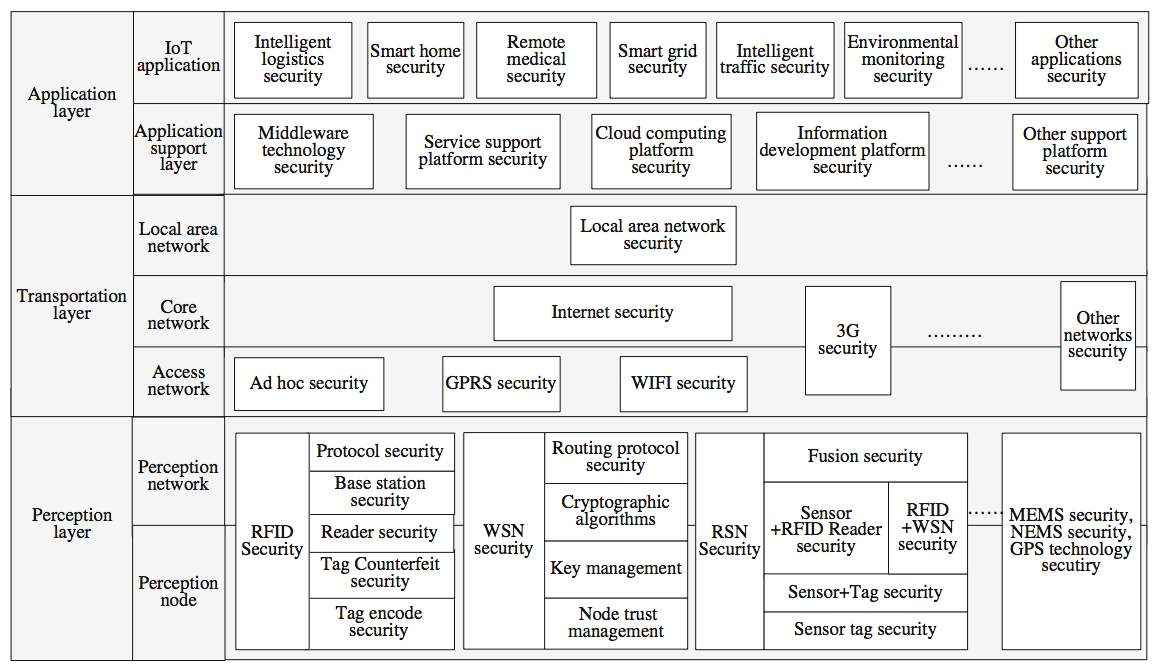}
\vspace{-1em}
\par\end{centering}
\caption{\label{fig:Security-issue-overview}Security issue overview suggested
by \cite{Jing:2014}}
\vspace{-1em}
\end{figure*}

From our analysis of the published papers in this direction, we recognized many findings and multiple challenges left to address for IoT and security crosscuts most of them. Looking into various published studies in the literature, we noticed that they concern security issues in three layers. Jing \textit{et al.} \cite{Jing:2014} summarized these layers as the perception layer, network layer, and application layer. Figure \ref{fig:Security-issue-overview} depicts the detail of this three-layered roadmap. 

The perception layer considers internal device security, such as particular sensor\textquoteright s concerns. The network layer looks at transmission, communication, and information security. Finally, the application layer involves the application--level perspective, such as service data security or security of support services. Some studies suggest to rather consider four layers in IoT, further dividing the application layer to an application and support layers, however, we consider both of these as a single layer.

There are various security perspectives we can consider delimited by the architectural layer, while a particular IoT design and features influence others. In the next subsections, we elaborate each layer in more detail.

\begin{itemize}
\item[\textbf{A. }] \textbf{\small{}Perception layer}{\small \par}
\end{itemize}

On the lowest perception layer, the security concerns address information collection and transmission. Since the \textquotedblleft  things\textquotedblright{} possess low computation power, they have limitations on complex security protections. At this layer, we recognized that most of the studies are dealing with things equipped with Radio Frequency Identification (RFID) and Wireless Sensor Networks (WSN).

For the RFIDs, we can summarize the following findings:
\begin{enumerate}
\item RFIDs must prevent exposition of private information. Thus tags cannot be read by everyone, hiding meta information such as labeling, chips, antenna details, etc. 
\item RFID signals need to be encrypted still preserving high-speed data
transmission, not impacting energy. 
\item Authentication and encryption allow both communicating peers to confirm the identity and preserve data confidentiality. 
\item Cryptography can help with privacy and confidentiality of the system. 
\item Side channel attack prevention should involve hiding or masking, to, e.g., prevent an attacker from exploring energy consumption and thus identifying internal details. 
\end{enumerate}
Regarding sensors, we must be aware that data are in the public space. An attacker can bring laptop and harvest all the data. Many studies found that we usually need to:
\begin{enumerate}
\item Deal with key management, in particular, key distribution and consequent management to keep them valid. 
\item Involve secret key algorithms to encrypt and decrypt messages. 
\item Secure routing for clustering, data fusion, multi-routing, etc. 
\item Integrate intrusion detection, monitoring the network and indicating
odds. 
\item Provide Authentication and Authorization control to provide node trust
\end{enumerate}
Various sorts of attacks have been found at this layer in the studies: 
\begin{enumerate}
\item A particular node can be captured and controlled by an attacker, leaking information or group communication, which threatens the entire subnet. 
\item A fake node can appear publishing false data; the receiver may get confused ignoring real data. 
\item This can lead to large processing on nodes draining battery charge leading into the fast discharge and network failure. 
\item DoS attack can target a particular processing node eventually bringing down entire infrastructure.
\item Timing attacks are analyzing the node processing of encryption to steal device key. 
\item Routing attacks are sending, tampering or re-sending routing information, possibly making loops impacting delay or even flooding the infrastructure. 
\item Replay attacks push fake messages, accepted in the past, to a processing node, expecting to gain to trust impacting certification and authentication. Side channel attack can be issued towards encryption devices. 
\item The side channel information can be exposed to the process of the device operation, such as time consumption, power consumption, etc.. 
\item Most of IoT solutions expect a uniform distribution of authentication. however, a mass node authentication can occur.
\end{enumerate}
\begin{itemize}
\item[\textbf{B. }] \textbf{Network layer (Transport)}
\end{itemize}
General networking security solutions address the middle layer. However still many possible attacks exist that we live with on nowadays Internet. Security problems can appear on communication network threatening data confidentiality and integrity. The common threats are illegal access, eavesdropping information, privacy damage, integrity damage, DoS attack, man--in--the-middle attack, virus invasion, and exploit
attacks.

This layer must also deal with the variety of sources and compatibility; existing security approaches emphasize human--computer interaction, while in IoT the attention it to machine--to--machine interaction. Support for various, heterogeneous endpoints only brings a high potential for errors and security vulnerability. Each device needs to be identified in the network, and the mass authentication can lead to congestion, with attacker potentially targeting the authentication server. Here, devices are often designed to harvest data and expose them; hackers can focus on retrieval and collect private information. 

Several studies suggested that this layer provides an environment for access, transmission, and store for the perception layer. It can be further divided into three sub-layers with specific functions, such as access, core, and local area networks. The access network deals with wireless connections (WiFi, Adhoc, GPRS, etc.). The core network is responsible for the data transmission. The local area network prevents data from leakage as well as applies server protection, e.g., involving network access control, denial of malicious code execution and removal of unused services. In fact, many studies found that DDoS is the most common sort of attacks in IoT. However, the threats found so far can be summarized as:

\begin{enumerate}
\item Unauthorized access 
\item Information theft or manipulation 
\item DDoS attacks 
\item Virus /Malware attack 
\item Scalability of authentication
\end{enumerate}
\begin{itemize}
\item[\textbf{C. }] \textbf{Application layer}
\end{itemize}
The particular domain strongly influences the top layer. For instance, in information systems, security authorizations are usually associated with roles, these can be context--sensitive based on particular data, location, time or combination of these. A particular service may require a given role to perform a given service. Specific users or applications granted a role based on their status or setting. In IoT generally, no standard exists and will differ for domains, such as smart homes/medical sensing systems, community, cities, etc. In general, this layer needs to recognize valid and spam data and filter them.

This layer may have an internal subdivision to application support layer that is usually more general, including middleware, machine--to--machine can be organized in different ways according to different services. Usually, it includes middleware, platforms, such as cloud computing and service support, etc. Since we expect big data, middleware addresses scalability and elasticity of services. The application layer most likely involves integration of business logic or the high--level system. It deals with privacy protection with fingerprinting, watermarking, etc.

Several studies found that at this layer we must deal with the following: 
\begin{enumerate}
\item Authentication to identify users 
\item Access permission to verify their right to perform an action 
\item Capability to deal with big data, as failed to scale leads into data
loss, and possible failure.
\item Heterogeneous system security concerns, for instance, buffer overflow.
\end{enumerate}

\subsubsection{Testing studies}
\label{subsec:testing_studies}

Area of studies focused directly on testing techniques and testing infrastructure is dominated by various reports on \textit{Security testing} approaches, techniques and frameworks (51 papers) and various \textit{Reports on test beds} created during previous IoT projects (47 papers). In the set of analyzed papers, we also identified another two related areas: \textit{Energy consumption testing} techniques and reports, discussed by six papers and \textit{Interoperability testing} techniques, being subject of seven relevant studies. We consider coverage of these two aspects relatively low due to their importance. Energy consumption aspect of the IoT devices has the direct impact on the reliability of the service, as well as on security aspects of the solution. Energy supply constraints might lead to the implementation of insufficient lightweight security algorithms and also to the impossibility to update IoT devices online, practically resulting in heterogeneity of variants of the devices deployed in production, causing combinatorial testing challenges later on during updates and maintenance of the system. Also, interoperability of the devices is one of the important aspects in the IoT systems, from the point of system reliability, as well as a possibility to integrate the particular solution with other IoT systems.

In the \textit{Security testing} area, several types of studies can be identified. The most of papers, 23, discuss particular \textit{Security analysis approaches and techniques}, spanning from various ethical--hacking techniques with the goal to detect a security flaw in an IoT system to approaches to detect security flaw based on analyses of collected system behavior data. Generally, these techniques can be based on \textit{Vulnerability models and metrics}, which are presented explicitly by seven analyzed studies. On top of these reports, technical implementation of particular \textit{Security analysis frameworks} has been described in 10 papers. Differently to cybersecurity testbeds classified as a separate category, these studies focus on technical frameworks to conduct the security testing process, rather than on particular configuration of the testbed to execute these tests. Finally, 11 studies presents various \textit{Security evaluation reports}, for instance in the Healthcare \cite{mcmahon2017assessing} or Sensor Networks \cite{asri2015impact} areas.

The most of the \textit{Reports on test beds} describe a general--purpose testbed for an IoT system, allowing End--to--End testing of this system (\textit{End--to--End testbeds}, 23 papers). However, also specialized testbeds for the network--level testing have been described in ten studies (category \textit{Network-focused testbeds}). Specifics of security testing leads to the construction of special \textit{Cybersecurity testbeds} for this purpose, and these projects have been described in five papers in the analyzed sample. Finally, a virtualization and cloud deployment trends can be clearly identified also in the sample of testbed reports (category \textit{Virtualized and cloud testbeds} represented by nine papers).

The IoT domain is considered as a source of testing challenges \cite{Marinissen2016,Foidl2016,Kiruthika2015} which we found in contradiction to the fact that ratio of the papers dedicated directly to the description of a particular testing technique specifically designed or adapted to IoT context is relatively small. Here, the research streams discussed above are the exceptions. 

Two possible scenarios can explain this situation. First, the IoT domain is not specific enough to justify the domain-specific testing technique. Second, the area of testing techniques that specifically designed for IoT solutions is significantly not covered in the current literature, and the definition of these techniques is pending as a future research task. 

Due to the analysis of the papers for this study and our previous experience and knowledge of the IoT domain, our subjective conclusion is that the former scenario is much more probable. Concerning current literature coverage of IoT-specific testing techniques, the following issues have been classified as significant from QA point of view:
\begin{enumerate}
\item Security issues 
\item Privacy issues
\item Performance issues 
\item Interoperability, missing or insufficient standards, proprietary standards vs. Internet standards 
\item Legislation issues 
\item Behavior of the system under a limited network connection 
\item Integration issues 
\item Number of various configurations and types of the end nodes, making the solution hard to test using all these combinations 
\item Focusing on test efforts efficiently to important aspects and critical
parts of the infrastructure regarding the security and privacy. 
\end{enumerate}

These topics are widely discussed in the related literature, as our mapping study show. Nevertheless, for the other aspects rated as important, we found rather little direct literature support (except the areas mentioned in this subsection).

Already existing testing techniques can cover some of the areas, and hypothetically, it is possible, that particular technique is not explicitly needed. As an example, we can discuss issue 8 in the above mentioned QA points. One can imagine current Constrained Interaction Testing (CIT) discipline \cite{ahmed2017constrained}, covering the problem. But would it be efficient not reflecting specifics of the IoT domain at all in the construction and application of a testing technique? For this issue, more efficient Feature Models as a test-basis for CIT, explicitly describing the IoT infrastructure can be created, including the modeling constructs added specifically to cover unique situations in IoT solutions. Also, as the variety of platforms and versions of IoT firmware makes the CIT problems extensive and here, new constraint solving techniques have to be invented, as the current techniques do not perform efficiently. 

Another example can be the issue 6 in the above mentioned QA points. Trivially, these situations can be covered by existing testing techniques for workflow testing (Process Cycle Test for instance \cite{bures2017prioritized}). However, such a process would be very probably sub-optimal. Extending an underlying model by a reliability meta-data allowing a simulation of node outage and redefinition of the technique to address this problem directly can be a typical example of the coverage we have in mind and which remains an inspiration for further research directions.

\subsection{Principal Testing Techniques Have Been Previously Studied (RQ4)}

Several areas can be tracked in the analyzed papers. Model-based testing as a primary principal research stream in software development quality assurance is also represented in the case of testing techniques discussed in IoT context. However, underlying models and process of test case generation differ. As a modeling layer, we can find examples of semantic description of IoT services \cite{kuemper2013test} or UML class and object diagrams combined with Object Constraint Language (OCL) \cite{Ahmad2016}.

The semantic description of IoT services and subsequent derivation of the test cases from this description proposed by Kuemper \textit{et al.} \cite{kuemper2013test} looks like a promising concept; however, issues may arise with keeping these semantic descriptions detailed enough and up to date with the SUT. More extensive automation of gathering these semantic descriptions from the actual state of the SUT would enhance the method further and might represent prospective future direction. The concept of MBT as a service for IoT platforms by Ahmad \textit{et al.} \cite{Ahmad2016} is relying on an established MBT approach based on OCL. The concept is valid, however, to achieve better efficiency, the MBT technique can be further extended to focus on the interoperability problem more systematically. One possible option is adding SUT configuration variants as an input into the MBT process pipeline presented in the study.

The Model checking is another principal testing technique, whose adoptions and applications are present in the analyzed sample. As underlying models, we can find formal specification languages \cite{Choe2016} or Computation Tree Logic (CTL) \cite{Jia2012}. 

A study by Choe \textit{et al.} \cite{Choe2016} proposes a modeling of a dynamic IoT system by a specialized formal specification language based on  $\delta$- Calculus and subsequent verification of dependencies among the movements in the IoT system using the Geo-Temporal Space (GTS) Logic. This proposal is primarily focused on modeling and verifying the dynamic properties of the systems, whose devices are mobile in a geographical environment. Despite the concept is promising, its verification on real examples is not presented in the study. Jia \textit{et al.} proposed model--checking approach for publish--subscribe systems, which is directly applicable also to IoT systems, as this domain overlaps with IoT. The system is modeled by CTL and the model can be generated from actual SUT code. However, information about the experimental verification of this concept is not satisfactorily provided in the paper.

Runtime verification as a related area had also its representative in the sample \cite{Torjusen:2014:TRV:2642803.2642807, Gonzalez10.1007/978-3-642-45364-9_26}. However, it seems that this area is rather emerging in the IoT context. 

A study by Torjusen \textit{et al.} \cite{Torjusen:2014:TRV:2642803.2642807} explores the possibility of runtime verification of adaptive security in IoT systems, however, the paper focuses specifically at eHealth applications and does not explicitly discuss possible extensions to other types of IoT systems and other application domains. Gonzalez \textit{et al.} \cite{Gonzalez10.1007/978-3-642-45364-9_26} propose runtime verification of behavior--aware mashups (web applications combining content from multiple sources and providing access to these sources via a unified user graphical interface), which can be considered as broader application scope in the IoT context. Unfortunately, broader experimental results reporting on the application of this proposal in an industry project are not presented in this study.

Model checking and Run-time verification, Reliability models, shall also be mentioned. Regarding this area, the first work related to IoT is focusing primarily on the combination of hardware and software \cite{AhmadIoT2014,Yong2014}.

Ahmad proposed a method to derive probabilistic estimates for the reliability of software and hardware in an IoT system \cite{AhmadIoT2014}. In his approach, he combines Numerical Finite Element Models (FEM), statistical techniques and Monte Carlo simulation. Unfortunately, the evaluation of this proposal is limited only to two relatively simple use cases from a telecommunication IoT system. Yong-Fei and Li-Quin proposed comprehensive evaluation model of the reliability of IoT systems. Proposed evaluation is based on the Analytic Hierarchy Process (AHP) method and might be promising for further applications, however, provided an experimental example presented in this study is rather limited. Analysis of behavioral patterns of the system on software level with the goal to improve the reliability of the system has been also explored by Shi \textit{et al.} \cite{Shi6805436}. As the paper is limited to the mote (a wireless transceiver also works as a sensor device) IoT applications, exploration of this approach in other IoT domains might represent a prospective future research area.

Also, Usability testing of IoT solutions can be identified as already covered by the first works, including the specific usability testing framework \cite{Wittstock2012}. In the related area, discussion on users' perception of IoT quality of service has been conducted \cite{shin2017conceptualizing}.

The usability testing framework by Wittstock \textit{et al.} \cite{Wittstock2012} aims at expressing underlying security in an IoT system to support user tests. In this support, virtual reality is used, which makes this concept innovative. The study focuses on smart home and smart office domains. Unfortunately, applicability on other domains of IoT systems (e.g. manufacturing or smart city) is not explicitly discussed in the paper. Shin examined Quality of Experience (QoE) of the users with the IoT system, namely the relation between system and data quality and subjective perception of IoT system by its users, satisfaction, and utilization of the system. Despite the study represents a valid approach to evaluate the user experience with the IoT system, its limit might be too general formulation of the questions answered by the users to evaluate the QoE.

Another research and development stream, which is significantly represented in the analyzed sample is related to the construction of efficient test environments (or test beds) for IoT solutions. Simulation of the devices is a logical option in this area for instance \cite{Palau2013}. In the analyzed samples, we can find examples of stand-alone tested setups \cite{Kawazoe2015}, distributed architectures \cite{Rosenkranz2015}, or crowd-sourcing based test beds \cite{Fernandes2015}. Also, the first testbed characterization works can be found, for instance, \cite{papadopoulos2017thorough}.

A study by Papadopoulos \textit{et al.} \cite{papadopoulos2017thorough} discussed contemporary testbed construction approaches in the area of RFID systems. The study analyzes and classifies the literature related to the RFID testbeds. However, a more detailed analysis of the literature is not presented. Instead, the authors focus on detailed analysis of particular FIT IoT-LAB testbed.

Regarding the focus on specific IoT solutions, examples of specifically-designed testing techniques can be found. For instance, the data-driven testing methodology for RFID systems presented by Lu \textit{et al.}\cite{Lu2010}. The proposed method is verified by mutation testing technique; however, a case study from a real industrial project would give more insight into the practical applicability of the method.

A particular area is protocol testing, which represents a significant part of the analyzed sample. The methods are varying here. For instance, statistical verification \cite{Bae2016}, formal verification \cite{Silva2016}, conformance testing \cite{Xie2013}, or randomness testing \cite{Gohring2015}.

Bae \textit{et al.} proposed a statistical verification of process conformance in an IoT system based on log equality test \cite{Bae2016}. The concept seems promising and is experimentally verified by an application on data from a system supporting steel manufacturing processes. To draw more conclusions about the efficiency of the method, more results from other application domains shall be provided. Silva \textit{et al.} \cite{Silva2016} proposed formal verification methods for cross--layer protocols used in IoT systems. However, the method is primarily designed for WSN systems, and the study does not discuss possible applicability to other IoT domains. Also, another protocol testing approach proposed by Xie at al. \cite{Xie2013}, based on conformance testing is focused primarily on WSN IOT systems. 
A study by Gohring and Schmitz proposes randomness testing approach for physical layer key agreement and show preliminarily promising results; however, verification of the method on larger test data sets shall be made.

\subsection{Specific Domains of IoT Systems Have Been Investigated from a Quality Viewpoint (RQ5)}

\begin{figure*}
\centering
\includegraphics[scale=0.6]{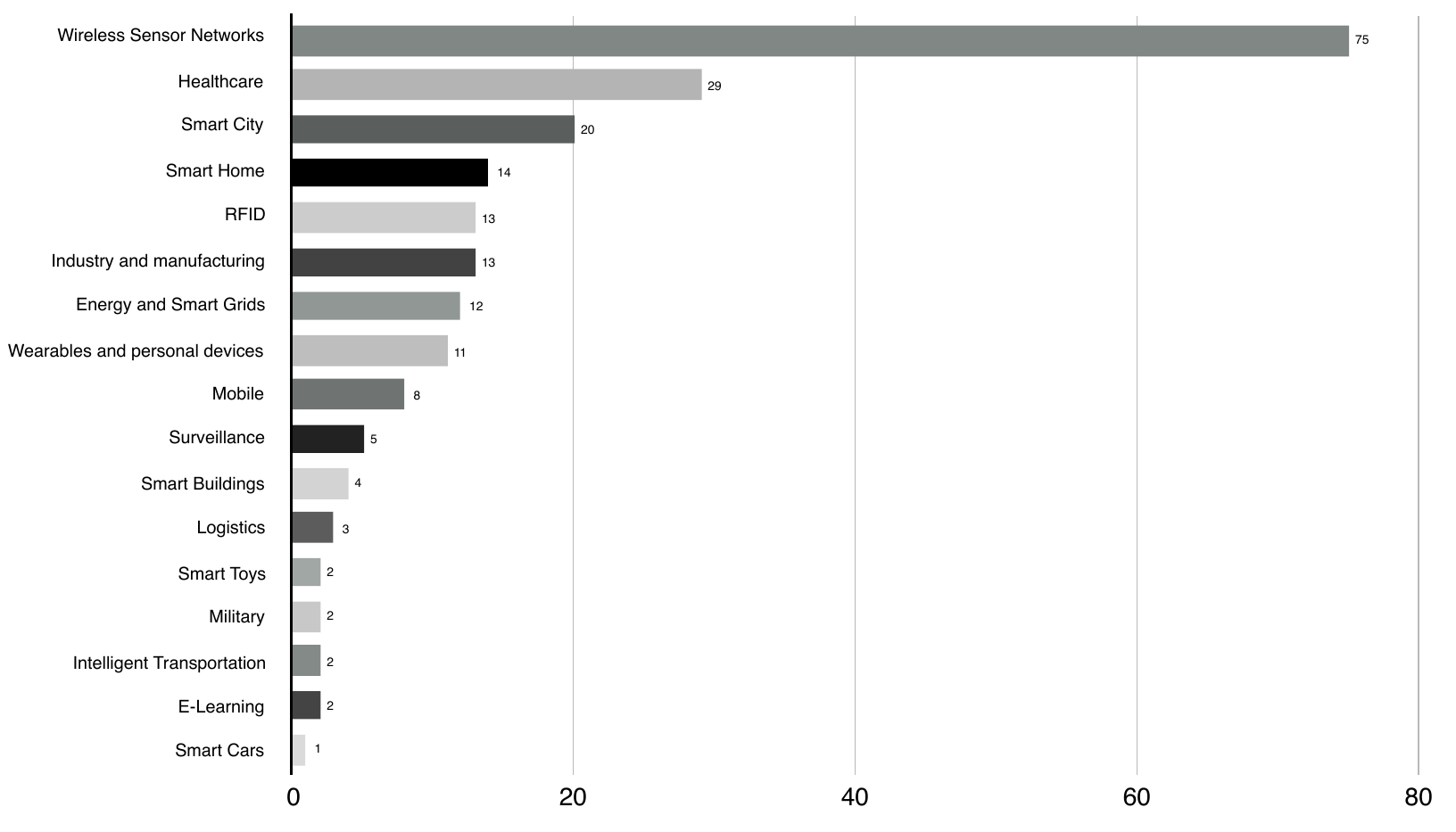}

\caption{\label{fig:application_domains}IoT application domains discussed in the analyzed papers}
\vspace{-1em}
\end{figure*}

To answer this research question, we investigated particular domains of IoT applications, which were discussed by the analyzed papers. Out of 478 studies, 216 papers were directly dedicated to particular IoT application domain. Figure \ref{fig:application_domains} presents an overview of these domains with respective numbers of papers.

Among the analyzed papers, \textit{Wireless Sensor Networks} is the IoT domain most frequently discussed from the quality viewpoint. From the analyzed sample, 75 papers are related to this area. Majority of these papers are discussing security issues (56 papers in total).  

Regarding the number of studies, the Wireless Sensor Networks domain is followed by \textit{Healthcare} systems discussed by 29 papers and \textit{Smart City} domain discussed by 20 papers. Also, Smart Home systems (14 papers), various RFID tracking systems (13 papers) and IoT--aided manufacturing systems (13 papers) have been discussed from the quality viewpoint quite frequently. 

Some areas are covered to marginally; this can be because the application area might be marginal from a research viewpoint, as \textit{Smart Toys} or \textit{E-learning}, or, research and development in the area is traditionally hindered from publication for security or competition reasons (\textit{Military} or \textit{Smart Cars}).

In the categorization presented in Figure \ref{fig:application_domains}, we made \textit{RFID} and \textit{Logistics} as two separate categories, as RFID systems can also be applied in other domains than logistics only. In the analysis of published studies, we also distinguished between Smart homes and Smart buildings. We understand \textit{Smart homes} as personal houses and apartments employing IoT devices to enhance the quality of living and security of these personal homes. In contrast, we use Smart buildings category for public or office buildings, benefiting from various IOT infrastructure to increase the quality of the workplace, secure conditions or to optimize the usage of these buildings.

We also distinguished \textit{Smart City} category from \textit{Intelligent Transportation} IoT--based systems. Intelligent Transportation focuses on optimization of general transport processes based on real--time data and this discipline spans beyond the IoT technology. The concept of Smart City is broader and employs the IoT technology to improve the quality of urban life and make the current cities more sustainable. IoT solutions, in which \textit{Mobile} devices and smartphones interacting with the system plays a central role are also given a special category in this analysis.

\subsection{Current Limitations and Challenges in QA for IoT (RQ6)}

In fact, we can find many limitations and challenges from addressing the RQ4. However, here, we can add more content to that discussion. Currently, IoT system designer usually deals with a dilemma whether to equip the solution with a lot of low-cost nodes (e.g., sensors), which deteriorate security or rather high-performance nodes capable of encryption or advanced features impacting the overall costs. One must be aware of lightweight mechanisms due to the limited performance of nodes/sensors, which gives preference to light encryption and authentication mechanisms that are easier to break. When designing one part of the system, we must consider that the underlying network is asymmetric. For instance, a network terminal is performant, while gateway nodes lack the performance. At the same time activities must be well coordinated no matter the endpoint, which also requires efficient endpoint management, usually involving assigning them keys. Since the IoT system applies to a variety of domains, one must remember that the particular context and domain strongly impacts the potential security concerns, further affecting any of the IoT layers.

In addition, there are two aspects of the IoT solutions, which can make research and development of proper quality assurance methods more challenging. The first is heterogeneity of IoT in general. Here, various types of solutions are produced (for instance sensor networks, smart home, intelligent transport, personal devices and much more), which could require specific testing methods. The second is the necessity to focus on various levels of the solution, including the physical layer, protocols, firmware and software of a particular device and end-to-end functionality from user\textquoteright s viewpoint. In this aspect, IoT differs from classical software testing, where we usually consider physical and protocol level as standardized and thoroughly tested already.

\section{Discussion\label{sec:Discussion}}

As can be readily observed from data, security and privacy aspects of IoT solutions are amply discussed and many alternative approaches proposed. These two aspects remain the primary challenge of IoT solutions, especially for devices, where ensuring of security aspects is the principal problem (for instance solar-powered devices with implied lightweight security algorithms or devices located in hardly accessible areas, where physical intrusion detection can be tricky). Also, limited software and firmware updates can contribute to this problem.

Besides security and privacy aspects, the relatively low number of papers is dedicated to specialized testing methods customized for IoT specifics. We can observe classical established testing research and development streams as model-based testing or model checking is applied to IoT domain, but, in contrast to the volume of the business potentially enabled by IoT, the number of relevant papers is surprisingly low.

Also, interoperability of the devices, protocols and a large number of their possible combinations to test shall, according to our opinion, be supported by more extensive research.

Development of quality assurance methods for IoT seems a bit reactive, following the technology development, which is enabling widespread and evolution of various IoT solutions. We consider this as a natural process, since a certain extent similar to the development of software quality assurance methods.

According to the state of the art, there are several areas, which are prospective for further research and development in IoT quality assurance methods. First, security and privacy issues shall be dealt with. Analyzed data show that a relatively high number of studies covers these topics. Nevertheless, the problem of security and privacy is still considered as not solved satisfactorily.

The next area is the interoperability of IoT devices. Here, we consider two streams as perspective: (1) IoT specific methods for integration testing and (2) methods how to combine efficient sets of device and infrastructure parts variants and versions, regarding heterogeneity of IoT solutions and sometimes even impossibility to upgrade or update to a newer version of device firmware or software. Adoption of Current combinatorial interaction techniques \cite{ahmed2017constrained} with the examination of their efficiency \cite{Bures2017} seems like a prospective way to solve problems arising from an enormous number of combinations of particular device variants.

The next area relates to a general testing strategy. For software projects, many guidelines on how to determine the intensity of testing for particular parts of the system under test and how to choose the best testing techniques to exist. The same shall be developed for IoT projects, respecting all specifics of IoT infrastructures.

Another area worth exploring is the development of specific test design techniques for testing of IoT solutions under limited network connection and related technical constraints. As users\textquoteleft{} dependency on Internet and IoT services grows continuously, this area is also becoming more relevant.

Finally, in the area of modeling of the system under test, suitable models for the semi-automated or automated generation of test cases shall be developed and verified in the practical model-based testing process. Differently, to classical software systems, these model shall also include physical and protocol layers, as these are much more heterogenic in the case of IoT solutions. 

\section{Threats to Validity\label{sec:Threats-to-Validity}}

Mapping study usually suffers from threats to validity. During our study, we have counted several threats that need to address. In fact, we have tried to eliminate the effect of these threats on the quality of the results and the outcome of the study. For these elimination activities, we decided to follow well--known methods to design our experiments. Some of those significant threats can be addressed here with our elimination mechanism. 

First, the selection of 100 percent related paper cannot be guaranteed although we have selected most of the papers that are within the scope of this study. We have tried to eliminate the effect of this threat by selecting and examining several search strings and conducting a pilot and snowball searches for several papers. 

The second potential threat to validity is the bias of the data extraction. The possible source of this bias could be one author extraction process when just one person extracts the information from the papers. We have also tried to eliminate the effect of this threat by distributing the data extraction among the authors and then each author double check other authors. Part of this elimination process is the use of automatic mining tools by spreadsheets for result verification by the data extraction process.

The third potential threat is the inclusion and exclusion of papers due to the scope of the paper. We have followed well-known methods for the selection criteria (see section \ref{subsec:Article-Selection-Criteria}). Owing to the broad range of papers that are dealing with the term IoT and the wide variety of published application using the concepts of IoT, we have spent much time in the scanning and reading the selected papers to assure that the papers are within the scope of the study. Hence, we have excluded those papers which are not related to the quality aspect of IoT. Those papers were covered by several studies related to other aspects of IoT such as \cite{Cavalcante2016,Gubbi2013,Atzori2010}.

The analyzed sample also does not include preprints of the papers submitted to or accepted in journals and conferences published by sites as \textit{arXiv.org}, \textit{researchgate.net} or on individual personal pages of the researchers active in the IoT domain. These preprints might contain novel ideas and quality assurance methods relevant to the scope of analyzed papers, however, to ensure objectivity and reliability of the information sources, we decided that the papers had to undergo the peer review process and had to be published by the journal or conference.

Another excluded set of paper is the set of those papers without specific output. We called those papers as \textquotedblleft opinion papers\textquotedblright{} which are just giving suggestions or opinions regarding IoT quality aspects but without experiments or robust proposed methods. The final set of excluded papers is the set of those papers which are not published in the considered academic databases. For the reliability of the results, we did not consider those papers which are published on unreliable sources on the Internet.

\section{Conclusions\label{sec:Conclusions}}

We have presented in this paper the results of mapping 478 published research papers related to different quality aspects of IoT. The mapping study takes the period between 2009 and 2017. We have gone through the detailed analysis of this population of papers from a different perspective based on a set of established significant RQs that have not been addressed before. In attempting to answering those RQs, we have gotten a set of significant results.

The results of the analysis showed us the dramatic increase in published research papers related to quality aspects of IoT. It appears from the results that majority of the papers published in conference and workshops while the rest were published in journals. We have highlighted those active researchers through their appearance as author/co--author of the published papers. We have further highlighted the active groups and countries. We have then arranged the contributions of the papers based on the quality aspects dealt with in the papers. We have given the detail of each study direction for those quality aspects. For each quality aspect, we have also discussed the methods and the principle techniques studied so far. We end this study with a discussion of limitations, challenges and areas for future research to improve the quality of IoT applications.

\section{Acknowledgment}

This study is conducted as a part of the project TACR TH02010296 ``Quality Assurance for Internet of Things Technology''.

\begin{table*}

\section*{Appendix A. List of Active Journals with Abbreviations}
\begin{centering}
\begin{tabular}{|l|l|}
\hline 
Acronym & Journal Full Name\tabularnewline
\hline 
\hline 
Futur Gener. Comp. Sy. & Future Generation Computer Systems\tabularnewline
\hline 
J. Netw.\& Comput. Appl. & Journal of Network and Computer Applications\tabularnewline
\hline 
Wireless. Pers. Commun. & Wireless Personal Communications\tabularnewline
\hline
IEEE IoT J. & IEEE Internet of Things Journal\tabularnewline
\hline 
Comput. Networks & Computer Networks\tabularnewline
\hline 
Ad Hoc Networks & Ad Hoc Networks\tabularnewline
\hline 
Comput. Electr. Eng. & Computers and Electrical Engineering\tabularnewline
\hline 
Comput. \& Secur. & Computers \& Security\tabularnewline
\hline
J. Ambient Intell. Hum. Comput. & Journal of Ambient Intelligence and Humanized Computing\tabularnewline
\hline 
Cluster Comp. & Cluster Computing\tabularnewline
\hline
Secur.\& Commun. Netw. & Security and Communication Networks\tabularnewline
\hline 
\end{tabular}
\par\end{centering}
\vspace{-.5em}
\end{table*}

\begin{table*}

\section*{Appendix B. List of Active Conferences with Abbreviations}
\centering{}%
\begin{tabular}{|l|l|}
\hline 
Acronym & Conference Full Name\tabularnewline
\hline 
\hline 
ICC & IEEE International Conference on Communications\tabularnewline
\hline 
ICC & IEEE International Conference on Communications \tabularnewline \hline
WF--IoT & IEEE World Forum on Internet of Things \tabularnewline \hline
IoT & International Conference on the Internet of Things \tabularnewline \hline
FiCloud & IEEE International Conference on Future Internet of Things and Cloud \tabularnewline \hline
Mobile Services & IEEE International Conference on Mobile Services \tabularnewline \hline
SecurIT & International Conference on Security of Internet of Things \tabularnewline \hline
HAS & International Conference on Human Aspects of Information Security, Privacy and Trust \tabularnewline \hline
ARES & International Conference on Availability, Reliability and Security \tabularnewline \hline
BODYNETS & International Conference on Body Area Networks \tabularnewline \hline
IMIS & International Conference on Innovative Mobile and Internet Services in Ubiquitous Computing \tabularnewline \hline
IoTS & International Internet of Things Summit \tabularnewline \hline
ICITST & International Conference for Internet Technology and Secured Transactions \tabularnewline \hline
IoTDI & IEEE/ACM International Conference on Internet--of--Things Design and Implementation \tabularnewline \hline
SpaCCS & International Conference on Security, Privacy and Anonymity in Computation, Communication and Storage \tabularnewline \hline
TRIDENTCOM & EAI International Conference on Testbeds and Research Infrastructures for the Development of Networks \& Communities \tabularnewline \hline
WoWMoM & IEEE International Symposium on a World of Wireless, Mobile and Multimedia Networks \tabularnewline \hline

\end{tabular}
\vspace{-1.3em}
\end{table*}

\bibliographystyle{IEEEtran}
\bibliography{sample-bibliography}
\end{document}